\documentclass[%
reprint,
superscriptaddress,
amsmath,amssymb,
prb,
floatfix,
longbibliography
]{revtex4-2}
\usepackage{graphicx}
\usepackage{amsmath}
\usepackage{amssymb}
\usepackage{mathtools}
\usepackage{xcolor}
\usepackage[colorlinks=true, citecolor=blue, urlcolor=blue, linkcolor=blue, bookmarks=false, hypertexnames=true]{hyperref}

\usepackage{array}
\usepackage[column=O]{cellspace}
\newcolumntype{P}[1]{>{\centering\arraybackslash}p{#1}}

\begin{document}
\graphicspath{}

\setlength{\parskip}{\medskipamount}

\title{Interacting holes in a gated WSe$_2$ quantum channel:\\ valley correlations and zigzag Wigner crystal}

\author{Jaros\l{}aw Paw\l{}owski}
\email{jaroslaw.pawlowski@pwr.edu.pl}
\affiliation{Institute of Theoretical Physics, Wroc\l{}aw University of Science and Technology, Wroc\l{}aw, Poland} 
\author{Daniel Miravet}
\affiliation{Department of Physics, University of Ottawa, Ottawa, Ontario, K1N 6N5}
\author{Maciej Bieniek}
\affiliation{Institute of Theoretical Physics, Wroc\l{}aw University of Science and Technology, Wroc\l{}aw, Poland} 
\author{Marek Korkusinski}
\affiliation{Quantum and Nanotechnologies Research Centre, National Research Council of Canada, Ottawa, Ontario, K1A 0R6}
\author{Justin Boddison-Chouinard}
\affiliation{Department of Physics, University of Ottawa, Ottawa, Ontario, K1N 6N5}
\affiliation{Quantum and Nanotechnologies Research Centre, National Research Council of Canada, Ottawa, Ontario, K1A 0R6}
\author{Louis Gaudreau}
\affiliation{Quantum and Nanotechnologies Research Centre, National Research Council of Canada, Ottawa, Ontario, K1A 0R6}
\author{Adina Luican-Mayer}
\affiliation{Department of Physics, University of Ottawa, Ottawa, Ontario, K1N 6N5}
\author{Pawel Hawrylak}
\affiliation{Department of Physics, University of Ottawa, Ottawa, Ontario, K1N 6N5}

\begin{abstract}
We present a theory of interacting valence holes in a gate-defined one-dimensional quantum channel in a single layer of a transition metal dichalcogenide material WSe$_2$. Based on a microscopic atomistic tight-binding model and Hartree-Fock and exact configuration-interaction tools we demonstrate the possibility of symmetry-broken valley polarized states for strongly interacting holes. The interplay between interactions,  perpendicular magnetic field, and the lateral confinement asymmetry together with the strong Rashba spin-orbit coupling present in WSe$_2$ material is analyzed, and its impact on valley polarization is discussed. For weaker interactions, an investigation of the pair correlation function reveals a valley-antiferromagnetic phase. For low hole densities, a formation of a zigzag Wigner crystal phase is predicted. The impact of various hole liquid phases on transport in a high mobility quasi-one dimensional channel is discussed.
\end{abstract}

\maketitle

\section{Introduction}
Monolayer semiconducting transition-metal dichalcogenides (TMDs) offer a unique platform for designing gated spin- and valleytronic quantum circuits due to efficient gate tunability of their electronic properties. This is because of the proximity of gates to two dimensional (2D) material, sizable band gap, highly controllable dielectric environment, as well as a strong spin–orbit coupling and the valley degree of freedom present in these materials~\cite{Goh2019,Sierra2021,Liu2019}.
Valley isospin can serve as a quantum information host in optically controlled devices~\cite{Xiao2012,Gong2013,Volmer2023} or gate-defined quantum dots (QDs)~\cite{pawlowski_2019, Altintas_Hawrylak_2021, pawlowski_2021,Rohling_2012,Burkard2023,Ciorga2000}.
Ongoing experiments focus on gated QDs in bilayer graphene~\cite{Allen2012,Eich2018,Banszerus2020,Tong2021}, monolayer TMDs~\cite{Song2015, Pisoni2018,Davari2020,Chouinard2021,Kumar2023}, or defect-defined TMD in-gap QDs~\cite{Krishnan2023}.
Monolayer 2D materials also serve as an efficient platform for studying many-body phenomena, including magnetism and correlations in graphene QDs~\cite{Potasz2009,Potasz2012}, fractional quantum Hall states in monolayer WSe$_2$~\cite{Shi2020} or bilayer TMDs~\cite{Reddy2023, zhao2023},  and valley-polarized states in WSe$_2$~\cite{Wang2017} and WS$_2$~\cite{Scrace_Hawrylak_2015}, 
as well as in gated QDs: MoS$_2$~\cite{Szulakowska_Hawrylak_2020}, WSe$_2$~\cite{Daniel2023} and bilayer graphene~\cite{Saleem2023}.
In order to develop quantum circuits based on spin and valley degrees of freedom in 2D materials, QDs and quantum point contacts, 1D channels, are necessary~\cite{Pisoni2017,Goossens2012b}. 

Inspired by a recent demonstration of a high-mobility one dimensional (1D) channel and the anomalous conductance quantization of holes in WSe$_2$~\cite{Justin2023}, also observed in bilayer graphene~\cite{Goossens2012b}, we present a theory of interacting holes in a 1D channel. We show that a possible explanation for the $e^{2}/h$ quantization~\cite{Justin2023} can be obtained only in regime of strong many-body hole-hole interactions~\cite{Scrace_Hawrylak_2015, Braz_Castro_2018, Donck_Peeters_2018}
and broken valley degeneracy.
Herein we discuss possible mechanisms and conditions required to lift the valley degeneracy.
We note that both the Hartree-Fock (HF) approximation and Quantum Monte Carlo (QMC) calculations~\cite{Attaccalite2002,Tanatar2008} predict lifting of spin degeneracy and spin polarization in a dilute 2D electron gas (2DEG). The anomaly in conductance of a 2DEG-based quantum point contact was attributed to spin polarization in a 1D channel~\cite{Wingreen2002a,Wingreen2002b}, although this was not confirmed by QMC calculations for a 1D channel~\cite{Guclu_Baranger_2009}. In TMD 2DEG, the effective mass HF predicts the spin and valley polarization~\cite{Scrace_Hawrylak_2015}. Results of  the exact diagonalization of a Hamiltonian of small number of interacting particles confined to a QD confirmed the existence of valley-polarized states~\cite{Szulakowska_Hawrylak_2020}. 

Coulomb interactions play an important role in our hole system forming a liquid phase. To characterize them, we introduce a characteristic length scale $r_s$ representing the average inter-hole distance in the liquid in units of Bohr radius $a_B\simeq0.529$~\AA. The ratio of the Coulomb interaction energy $V/\epsilon$, which scales as $r_s^{-1}$, to the kinetic energy $T$, scaling as $r_s^{-2}$, is proportional to $r_s$~\cite{giuliani2008quantum}. The estimated $r_s$ distance for the analyzed here hole channel is about $\sim{}40$.
In case of a 2D interacting electron liquid, at high densities ($r_s<25$) the ground state (GS) is unpolarized (paramagnetic), whereas for smaller densities ($r_s>25$) the ferromagnetic phase becomes lower in energy, giving rise to polarised GS \cite{Attaccalite2002}. Further lowering of density  leads to  Wigner crystallization.
We study here the possibility of valley polarized states in 1D channel both in HF and using exact diagonalization tools including pair correlation function (PCF) analysis. We do find partially valley-polarized states as well as  valley antiferromagnetic phases and zigzag Wigner crystal states~\cite{Matveev_2004}.

The paper is organized as follows:
We start with an introduction, followed by a single-particle model of a 1D channel.
Next, we explore the effect of many-body interactions on the ground and excited states of up to six holes in the channel, comparing HF with configuration-interaction results. We examine the nature of the partially spin and valley polarized ground state. Next, we introduce the pair correlation function, and predict the emergence of valley antiferromagnetic phases and Wigner crystallization in the channel. We finish with conclusions and an outlook.
\section{Single-particle states}
In order to determine the role of hole-hole interactions in the valley polarization we first develop a single-particle model of a 1D channel following Boddison-Chouinard \textit{et al.} [\onlinecite{Justin2023}]. To calculate single-hole states we use a tight-binding model for the WSe$_2$ monolayer~\cite{Bieniek_Hawrylak_2018, Bieniek_Hawrylak_2020, Szulakowska_Hawrylak_2020}.
Subsequently, we construct many-body configurations using the basis of the single-hole states.

\begin{figure}[t]
    \centering
    \includegraphics[width=.48\textwidth]{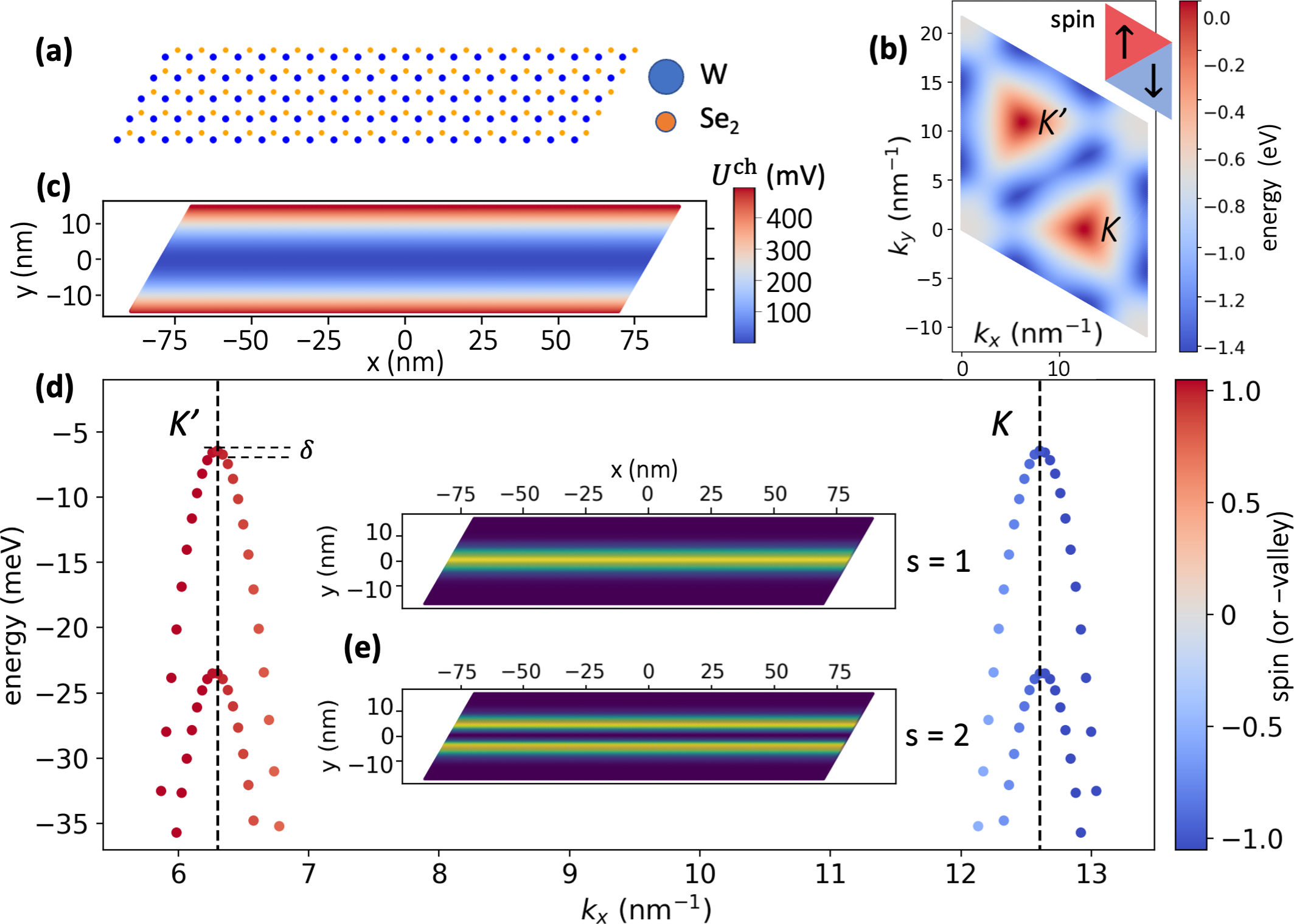}
    \caption{\label{fig:1}(a) Computational rhombus representing WSe$_2$ monolayer. 
    (b) Valence band energy surface within the Brillouin zone with two maxima at $K'$ and $K$ points. (c) Gate-defined channel potential $U^\mathrm{ch}$ applied along the $x$-axis. (d) Single-particle states calculated with the tight-binding model for WSe$_2$ monolayer with applied potential $U^\mathrm{ch}$. The hole eigenstates characterized by wave vector $k_x$ form parabolic subbands located in $K'$ valley with spin-up (red dots), or $K$ valley with spin-down (blue). These subbands are numbered by $s$ with densities (e) resembling harmonic oscillator modes.}
\end{figure}
A single-particle channel wavefunction for a hole state~$i$ satisfies the Schr\"odinger equation \cite{Altintas_Hawrylak_2021, Bieniek_Hawrylak_2020}:
\begin{equation}\label{eq:schroedinger}
\left(H_\mathrm{bulk}+H_{SO}+|e|U^\mathrm{ch}\right)|\Psi^i\rangle=E^i|\Psi^i\rangle,
\end{equation}
where the channel is defined within a WSe$_2$ monolayer lattice and oriented along the $x$-axis of an elongated, rhombohedral computational box presented in Fig.~\ref{fig:1}(a). The 1D electronic confinement is determined by an applied external gate-defined potential $U^\mathrm{ch}$ shown in Fig.~\ref{fig:1}(c). A hole is free to move in the channel direction, while its motion in the perpendicular direction is restricted by a Gaussian-like potential along the $y$-axis
$U^\mathrm{ch}(y)=U_0[1-\exp\!{\big(-\frac{y^2}{2 w^2}\big)}]$,
parametrized by the potential depth $U_0$ and width $w$.

We start by wrapping the computational rhomboid (in our case with size: $N_1=480$, $N_2=120$) onto a torus and impose periodic boundary conditions. This approach yields a set of allowed, discrete vectors $\mathbf{k}$ within a rhombus shown in Fig~\ref{fig:1}(b). 
The valence band (VB) wavefunction for each $\mathbf{k}$ is a linear combination of Bloch functions on the $W$ and Se$_2$ dimer sublattices $l$:
$|\psi_{\mathbf{k},\sigma}^\mathrm{VB}\rangle=\sum_{l=1}^{11} A_{\mathbf{k},\sigma,l}^\mathrm{VB} \left|\psi_{\mathbf{k},l}\right\rangle\otimes\left|\chi_\sigma\right\rangle$,
where $\left|\chi_\sigma\right\rangle$ is the spinor part of the wavefunction, and
$\left|\psi_{\mathbf{k},l}\right\rangle=\frac{1}{\sqrt{N_c}}\sum_{R_l}^{N_c} e^{i\mathbf{k}\cdot\mathbf{R}_l}\varphi_l(\mathbf{r}-\mathbf{R}_l)$
are Bloch functions built from atomic orbitals $\varphi_l$: five $d$-orbitals for tungsten and six $p$-orbitals (3 even and 3 odd with respect to metal-plane reflections) localized on Se$_2$ dimers.
The total number of unit cells is given by $N_c=N_1 N_2$, while $\mathbf{R}_l$ defines the position of atomic orbitals in the computational rhomboid.
To capture the strong spin-valley locking mechanism for holes in TMD's, the model includes an intrinsic spin-orbit interaction (SOI)~\cite{Roldan2014,Altintas_Hawrylak_2021, Bieniek_Hawrylak_2020}:
$H^{ll'}_\mathrm{SO}=\sum_a\frac{\lambda_a}{\hbar}\langle\varphi_l|\hat{\mathbf{L}}\cdot\hat{\mathbf{S}}|\varphi_{l'}\rangle$, $a=\mathrm{W},\mathrm{Se}$.
By diagonalizing the $22\times22$ Hamiltonian 
$H_\mathrm{bulk}+H_\mathrm{SO}$ 
at each allowed value of $\mathbf{k}$ we obtain the bulk
energy bands $E_{\mathbf{k},\sigma}^\mathrm{VB}$ and wavefunctions $A_{\mathbf{k},\sigma,l}^\mathrm{VB}$. 
The topmost VB energy surface is shown in Fig.~\ref{fig:1}(b), where two non-equivalent maxima corresponding to $K$ and $K'$ valleys, are clearly visible (inset showing the spin texture).
The structural and the tight-binding Slater-Koster parameters are listed in the Appendix, Section~A. 

In the next step, we expand the valence hole channel wavefunction $|\Psi^i\rangle$ in terms of the lowest energy valence Bloch states $|\psi_{\mathbf{k},\sigma}^\mathrm{VB}\rangle$:
\begin{equation}\label{eq:singehole}
|\Psi^{s,k_1,\sigma}\rangle=\sum_{k_2}B^{\mathrm{VB},s}_{k_1,k_2,\sigma}\left|\psi_{k_1,k_2,\sigma}^\mathrm{VB}\right\rangle,
\end{equation}
where $k_1$ and $k_2$ are indices numbering allowed values of the vector $\mathbf{k}$, i.e., $\mathbf{k}=k_1\mathbf{b}_1/N_1+k_2\mathbf{b}_2/N_2$, 
and
$\mathbf{b}_1=\frac{2\pi}{a}(1,-\frac{1}{\sqrt{3}})$, $\mathbf{b}_2=\frac{2\pi}{a}(0,\frac{2}{\sqrt{3}})$ 
are the reciprocal lattice vectors.
Finally, we solve the Schr\"odinger Eq.~(\ref{eq:schroedinger}) with the potential term $U^\mathrm{ch}$ by converting it to an integral equation for the coefficients $B^{{\mathrm{VB},s}}_{k_1,k_2\sigma}$,
\begin{equation}\label{eq:integral}
E^\mathrm{VB}_{k_1,q_2,\sigma}B^{\mathrm{VB},s}_{k_1,q_2,\sigma}+\sum_{k_2}B^{\mathrm{VB},s}_{k_1,k_2,\sigma}V_{k_1,q_2;k_1,k_2,\sigma}^\mathrm{ch}=E^s_{k_1\sigma}B^{\mathrm{VB},s}_{k_1,q_2,\sigma},
\end{equation}
where the electrostatic potential $U^\mathrm{ch}$, shown in Fig.~\ref{fig:1}(b), dressed by the band contribution $A^\mathrm{VB}_{k_1,k_2,\sigma,l}$ leads to scattering elements:
\begin{equation}
V_{k_1,q_2;k_1,k_2,\sigma}^\mathrm{ch}=\sum_l A^\mathrm{VB,\ast}_{k_1,q_2,\sigma,l}A^\mathrm{VB}_{k_1,k_2,\sigma,l}\,U_{l}^\mathrm{ch}(k_1,q_2;k_1,k_2).\nonumber
\end{equation}
Here $U_l^\mathrm{ch}(k_1,q_2;k_1,k_2)$ 
is the Fourier transform of the channel confinement on each sublattice $l$:
$U_l^\mathrm{ch}(\mathbf{q},\mathbf{k})=\frac{1}{N_c}\sum_{R_l}^{N_c}
U^\mathrm{ch}(\mathbf{R}_l)e^{i(\mathbf{k}-\mathbf{q})\cdot\mathbf{R}_l}.$
We note that $U^\mathrm{ch}$ does not mix states with different $k_1$, i.e. $U_l^\mathrm{ch}(q_1,q_2;k_1,k_2)=U_l^\mathrm{ch}(k_1,q_2;k_1,k_2)\delta_{k_1,q_1}$,
and its dependence on sublattice $l$ can be safely neglected: $U^\mathrm{ch}_l=U^\mathrm{ch}$.

By solving Eq.~(\ref{eq:integral}) we obtain the single-hole channel states $|\Psi^{s,k_1,\sigma}\rangle$ with energies $E^s_{k_1,\sigma}$, shown in Fig.~\ref{fig:1}(d) for $U_0=1.5$~eV and $w=30$~nm. They are characterized by the subband index $s$, wave vector $k_1$ in the channel direction, and spin $\sigma$.
They belong to two different valleys: $K'$ with spin-up and $K$ with spin-down, creating degenerate spin-valley-locked states -- see Fig.~\ref{fig:1}(d). Characteristic parabolic subbands are formed due to the lateral quantization in the channel potential $U^\mathrm{ch}$,  labeled by different values of the quantum subband number $s$. The two lowest energy subband wavefunctions are shown in Fig.~\ref{fig:1}(e). The second spin-valley pair (not shown) is split-off by $0.5$~eV by the strong intrinsic spin-orbit coupling in the VB. 

\section{Many-body effects}
We now turn to discussing the hole-hole interactions. We start with the basis of single-hole states
$|i\rangle=|\Psi^{s,k_1,\sigma}\rangle$.
In this basis the hole-hole Hamiltonian can be written as:
\begin{equation}\label{eq:Hci}
H_{hh}=-\sum_{i}E^i h^\dag_i h_i+\frac{1}{2}\sum_{ijkl} \langle{}ij|V/\epsilon|kl\rangle h^\dag_i h^\dag_j h_l h_k,
\end{equation}
where $h^\dag_i$ creates a hole in a single-particle state $|i\rangle$, $\langle{}ij|V/\epsilon|kl\rangle\equiv{}V_{ijkl}$ are the screened Coulomb matrix elements, and 
$\epsilon$ is the dielectric constant. 
The $V_{ijkl}$ elements are detailed in the Appendix, Section~C. 
One should note that the many-body model (\ref{eq:Hci}) is defined in the basis of the channel single hole states (\ref{eq:singehole}) being spatial compositions of Bloch states (forming “envelopes”), rather than from localized on-site Wannier-like orbitals.

We first define a basis of a finite number of single-hole states $|i\rangle$, fill this set of single-particle states with a given number of holes $N$, and generate all possible configurations: 
$|p\rangle=h^\dag_{i_1} h^\dag_{i_2}\dots h^\dag_{i_N}|0\rangle$.
We next expand the many-body wavefunction  in all possible configurations $p$ with coefficients $A_p^\nu$:
$\Psi^{\nu} = \sum_p A_{p}^{\nu}|p\rangle$,
with $A_{p}^{\nu}$ and $E_{\nu}$ being, respectively, the eigenvectors and eigenvalues of the correlated many-hole state $\nu$.
To generate these correlated states we turn on the interactions between configurations induced by the Coulomb interaction and build the configuration-interaction (CI) Hamiltonian $H_{hh}$ Eq.~(\ref{eq:Hci}). 
We diagonalize it in the space of configurations and analyze many-body spectra for different strengths of Coulomb interaction.
In particular, we calculate the total valley polarization 
$P=\sum_{p}|A_p^{\nu}|^2\left(\langle{}p|K'|p\rangle-\langle{}p|K|p\rangle\right)$, with $\langle{}p|K'|p\rangle$ counting holes in the  valley $K'$ in  configuration $p$ (and the same for valley $K$).
Due to the spin-valley locking the valley polarization is equivalent to the $z$-component of the total spin $S_z$.
In these calculations we assume $N=6$ holes, a minimal number to “symmetrically” populate the two valleys and states within each valley.

\subsection{Hartree-Fock states}
We start with an analysis of different configurations under the simplified HF approximation.
It differs from the exact CI method in that the many-body state  in the HF approximation is composted of only a single configuration of a Slater determinant form.
Thus, in contrast to CI where the many-body state is constructed as a superposition of multiple configurations (Fock states), in HF we compare the energies of each configuration separately.
While the standard (restricted) HF method cannot describe, e.g., Wigner crystallization, its extension called unrestricted HF (UHF) could~\cite{Goldberg2024}.

\begin{figure}[tb]
    \centering
    \includegraphics[width=.49\textwidth]{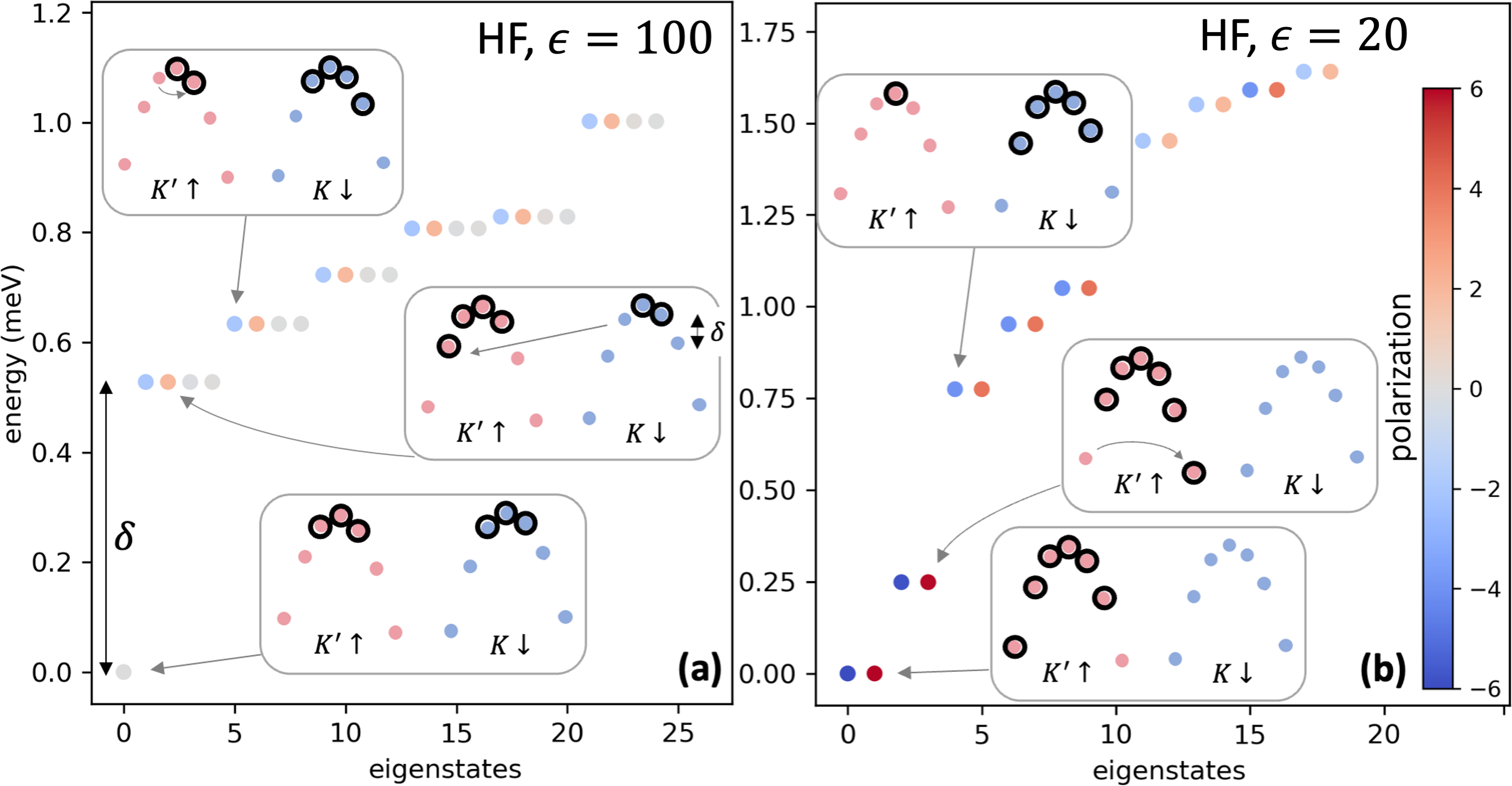}
    \caption{\label{fig:hf} Energies of Hartree-Fock configurations of $N=6$ holes in the channel: (a) With no interactions, for $\epsilon=100$, the ground state is unpolarized and separated from the polarized states by the single particle gap $\delta$ resulting from the channel finite-size. (b) For finite interactions, controlled by $\epsilon=20$, the ground state is strictly valley-spin polarized. Insets: black circles around single-particle states in both valleys denote states occupied in a given configuration.}
\end{figure}
Fig.~\ref{fig:hf}(a) illustrates
configurations and their energies for a noninteracting, strongly screened channel with $\epsilon=100$. The lowest-energy configuration consists of 3 holes at the top of the $K$ valley with spin-down and 3 holes at the top of the $K'$ valley with spin-up. With an equal number of holes in each valley there is no valley and spin polarization and the color of this configuration with the reference energy $E=0$ is gray.  The next configuration involves an excitation corresponding to the transfer of a hole from the valley $K$ to the valley $K'$, introducing a degree of valley and spin polarization. More configurations are constructed by occupying low-energy states of the 1D channel, and their color reflects the degree of their valley polarization. In Fig.~\ref{fig:hf}(b) we show the energy of HF configurations for interacting valence holes (with lowered $\epsilon=20$). We see that the lowest-energy configurations correspond to fully valley polarized states, with 6 holes occupying lowest single-particle states of either valley $K'$ (red) or $K$ (blue). Intense red (blue) color for these polarized configurations reflects full valley (and spin) polarization. The energy of polarized configuration is low because in a single valley, holes lower their energy by intravalley exchange, even though they have to increase their energy by occupying the ladder of subsequent single-particle energy states in the given valley~\cite{Szulakowska_Hawrylak_2020}.

\subsection{Correlated states}
We next move on to the CI method and diagonalize $H_{hh}$ (Eq.~\ref{eq:Hci}) in the configuration space and obtain correlated hole states as linear superpositions of many configurations. 
In the presented calculations $\binom{24}{N}=134596$ configurations were constructed. The description of the CI method as well as details on the calculation of the Coulomb matrix elements are given in the Appendix, Section~C. 

\begin{figure}[tb]
    \centering
    \includegraphics[width=.49\textwidth]{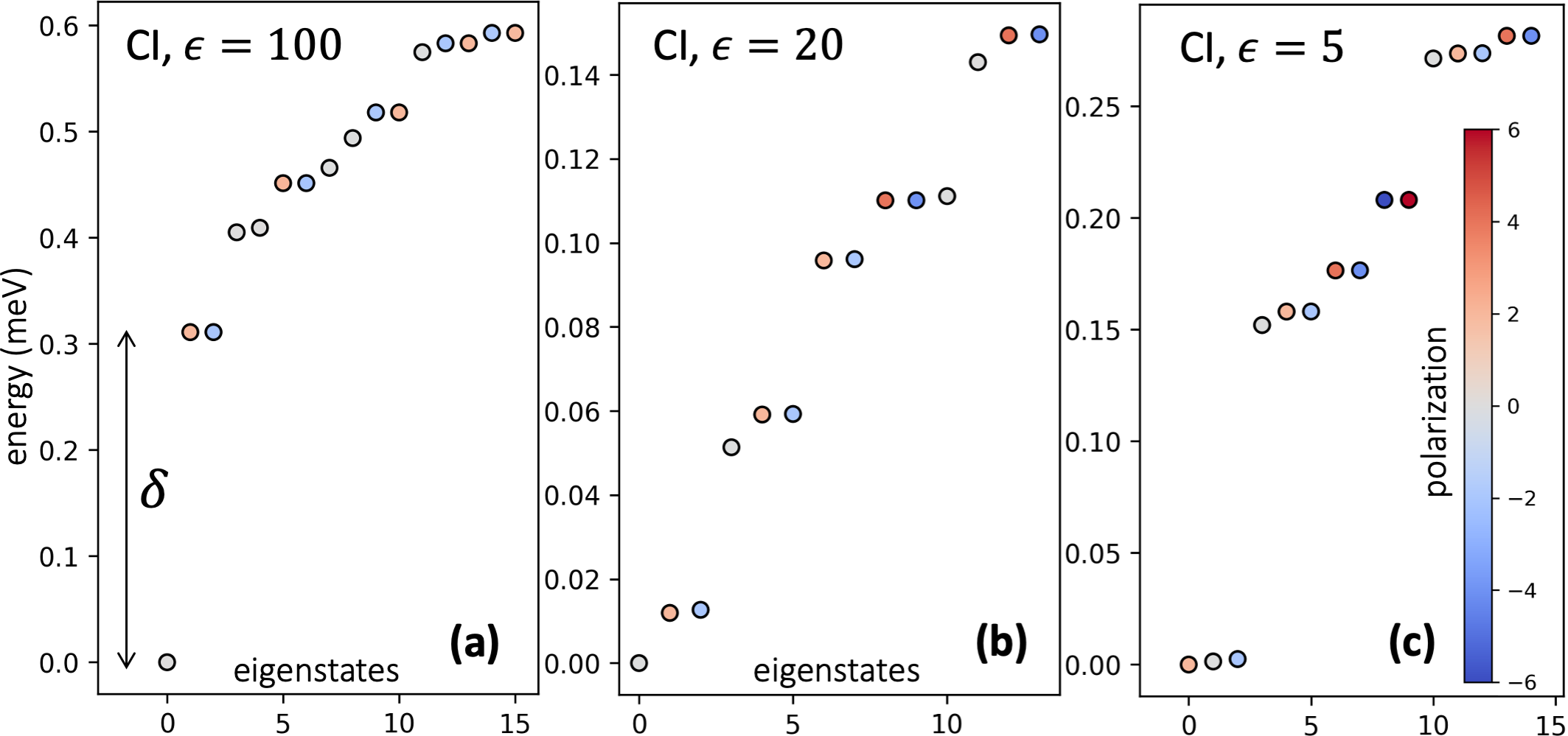}
    \caption{\label{fig:ci}Energies of correlated states of $N=6$ holes for different interaction strengths (decreasing $\epsilon$). (a) Single-particle energy gap $\delta$ due to the finite length of the channel is reduced by correlations giving a degenerate triplet (c), resulting in partially polarized ground state.}
\end{figure}
Fig.~\ref{fig:ci} shows the low-energy spectrum $E_{\nu}$ of correlated holes for different strengths of the Coulomb interaction, controlled by the dielectric constant $\epsilon$. The spin-valley polarization is marked by colors as indicated. 
Valley-spin polarized configurations are  doublets, marked here with red and blue colors, while unpolarized configurations are marked in grey. Spectra for different strengths of Coulomb interactions are shown.
For weak interactions (Fig.~\ref{fig:ci}(a) with $\epsilon=100$), the ground state  is valley unpolarized and separated from the rest of the spectrum by a single-particle gap $\delta$ between the filled and empty states.
Increasing interactions reduces the gap and eventually, for strong interactions $\epsilon\lesssim{}5$, shown in Fig.~\ref{fig:ci}(c), the characteristic triplet forms.
The three degenerate states, representing the valley triplet, form a GS with partial polarization.
We note that the HF results differ from the correlated ones with partial valley polarization, also observed for WSe$_2$ gated QDs~\cite{Daniel2023}.
HF approximation is known to overestimate the valley/spin polarization~\cite{giuliani2008quantum}. That is why when analyzing polarized states and symmetry breaking phenomena HF approximation might be insufficient.
The emergence of the valley triplet GS can also be understood using simple analytical considerations including time-reversal symmetry (TRS), for such discussion see Appendix, Section~B.

\subsection{Effect of the magnetic field and the Rashba SOI on ground state}
\begin{figure}[bt]
    \centering
    \includegraphics[width=.48\textwidth]{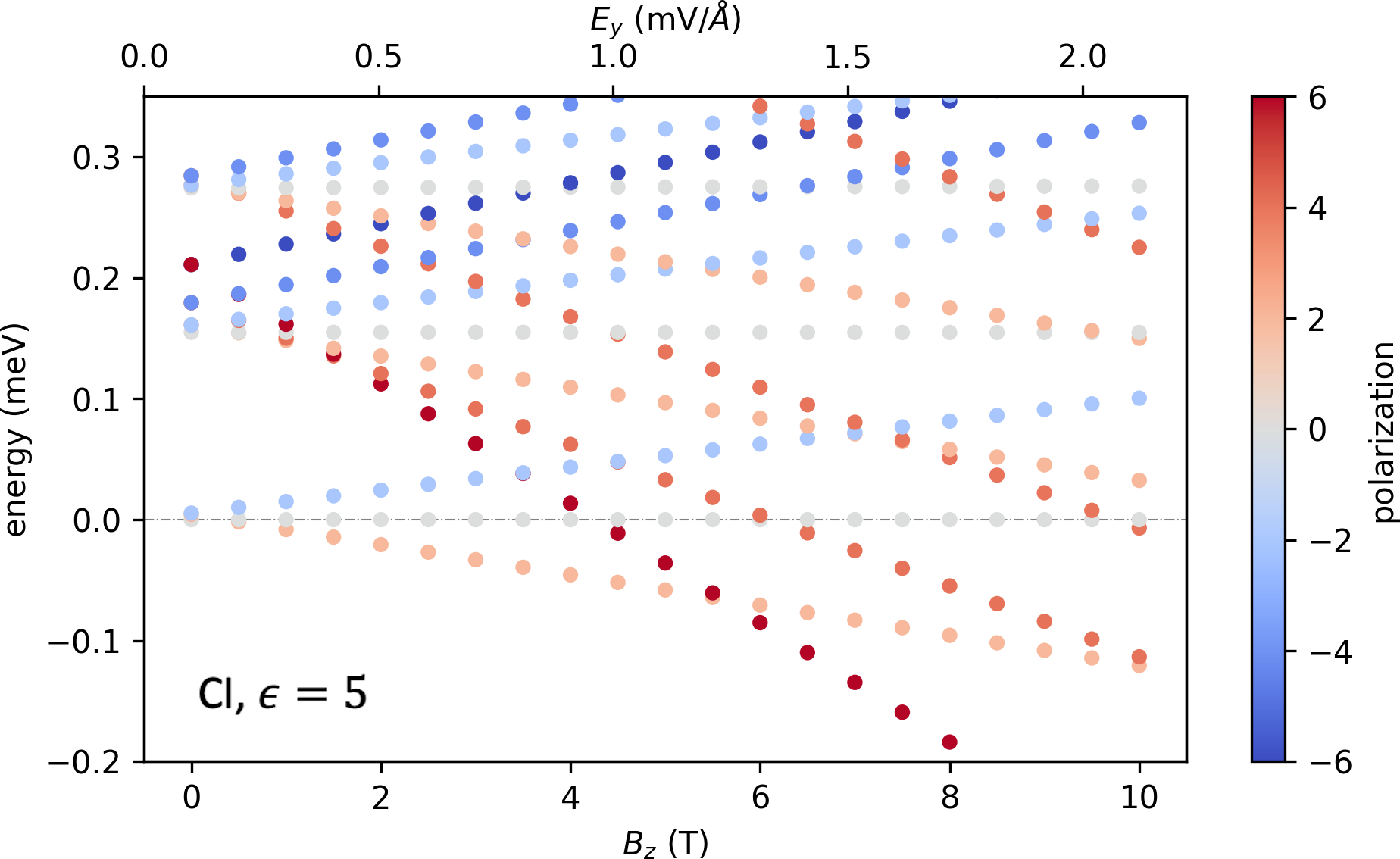}
    \caption{\label{fig:zeeman} Spin/valley splitting of degenerate GS induced by a perpendicular magnetic field $B_z$, bottom x-axis, or lateral confinement asymmetry $E_y$, top x-axis. For clarity, the energy offset was added so that the lowest unpolarized state is always shifted to zero. This calculation assumes $\epsilon=5$.}
\end{figure}

The simplest way to split polarized degenerate states in the GS valley triplet is to apply a perpendicular magnetic field $B_z$. The Zeeman term $H_Z=g\mu_B B_z\sigma_z/2$, with assumed spin-valley g-factor $g\simeq10$, yields spin/valley splittings as presented in Fig.~\ref{fig:zeeman}.

Another approach to create a spin/valley splitting involves lateral structure inversion asymmetry which is manifested by the lateral electric field $E_y$. 
This can lead to the Rashba SOI field of the form~\cite{Winkler,pawlowski_2016}: $H_R=\alpha_R|e|\langle E_y\rangle p_x\sigma_z/\hbar$, that generates an energy splitting even though the Rashba coupling fulfils TRS. For a realistic lateral field of $E_y\simeq1$~mV/\AA, the hole velocity during the transport measurements upon source-drain bias  of $10^5$~m/s, $m=0.35\,m_e$, and the Rashba coupling parameter for WSe$_2$~\cite{Kormanyos} $\alpha_R=0.18$~\AA$^2$, gives  an energy splitting for opposite spins 
 of $0.054$~meV -- see energy levels vs. top x-axis in Fig.~\ref{fig:zeeman}.
Note also that  the single-particle eigenstates dressed by the Rashba interaction acquire a plane wave component  $e^{\langle\sigma_z \rangle i q x}$, with wavevector $q=m\alpha_R|e|\langle E_y\rangle/\hbar^2$~\cite{Catstates}. This, however, does not change the Coulomb matrix elements. 

\section{Pair correlation function}
In the following section we discuss the spatial dependence of hole density and the spatial hole correlations by analysing a spin-resolved pair correlation function (PCF).
\subsection{Valley AFM ordering}
\begin{figure}[bt]
    \centering
    \includegraphics[width=.45\textwidth]{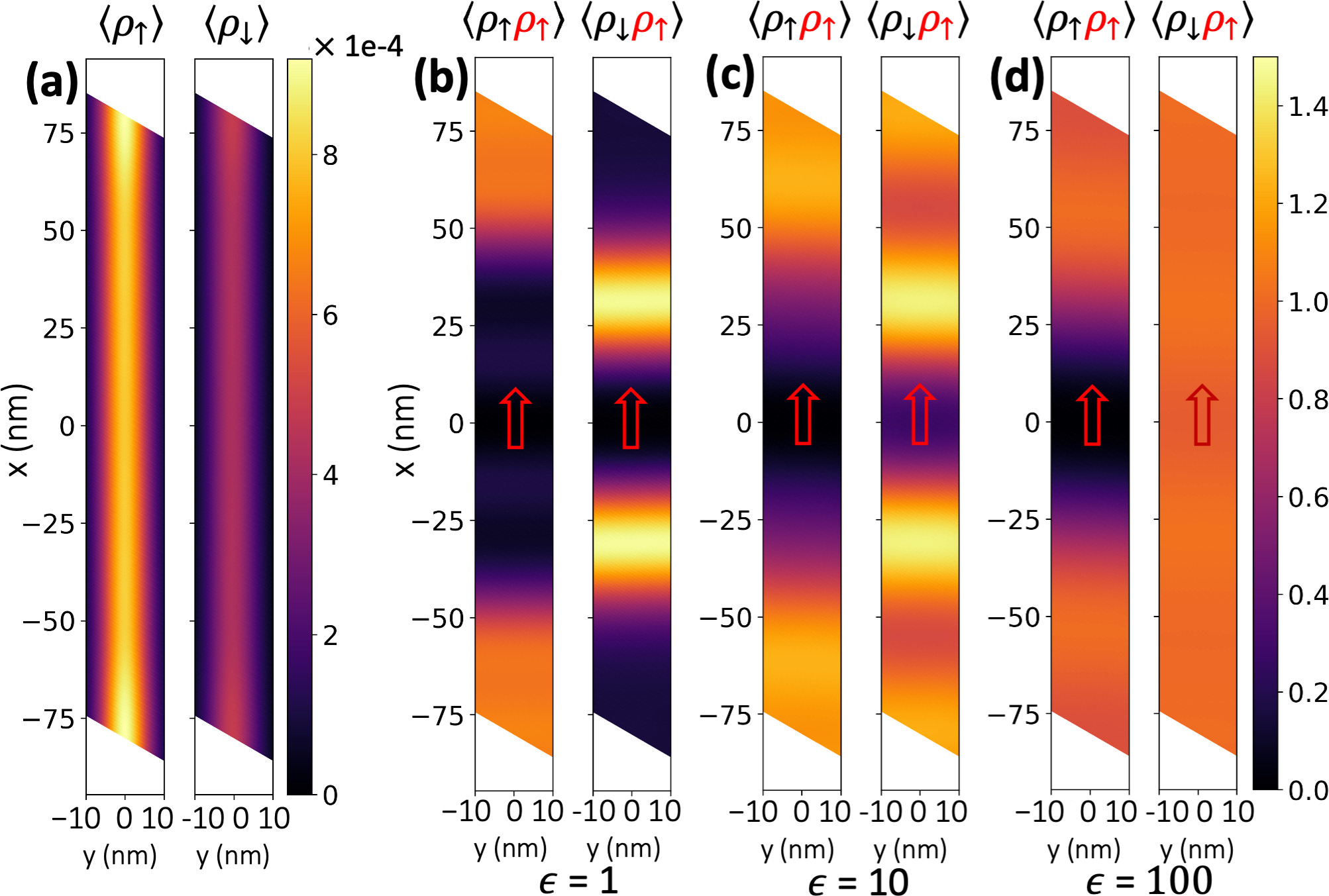}
    \caption{\label{fig:pair} Liquid of $N=6$ holes formed in the 1D channel of width $w=30$~nm: (a) hole spin-density, (b-d) spin pair correlation plotted for different Coulomb interaction strengths. The characteristic dip around $\mathbf{r}=(0,0)$ for the up-up component is known as the exchange hole. (c) For a dielectric constant of  $\epsilon>5$  we observe the characteristic spin-valley AFM ordering. (d) For weak interactions the down-up component becomes flat.}
\end{figure}
Let us now analyze different phases of the hole liquid in the 1D channel. 
In Fig.~\ref{fig:pair} we present plots of the spin-resolved PCF for the hole liquid composed of $N=6$ holes with one spin-up hole located at the position $\mathbf{r}_0=(0,0)$, marked as the red arrow in Figs.~\ref{fig:pair} and \ref{fig:wigner}. 
The spin resolved PCF for a correlated hole state $\Psi^\nu$ is defined as (for more details see Appendix, Section~D):
\begin{align}
&\langle\hat{\rho}_{\sigma}(\mathbf{r})\hat{\rho}_{\sigma'}(\mathbf{r}_0)\rangle=\\
&=\!\frac{1}{2}\sum_{p,q}A^{\nu\ast}_p A^\nu_q \!\sum_{i,j,k,l} \langle p| h^\dag_i h^\dag_j h_l h_k|q\rangle
\langle i|\hat{\rho}_{\sigma}(\mathbf{r})|k\rangle
\langle j|\hat{\rho}_{\sigma'}(\mathbf{r}_0)|l\rangle
\nonumber,
\end{align}
with $\langle i|\hat{\rho}_{\sigma}(\mathbf{r})|j\rangle=\Psi^\dag_i(\mathbf{r})\Psi_j(\mathbf{r})$ being the single-hole density operator elements. 
The calculated PCF behaves as follows~\cite{Rapisarda1996,Ortiz1994,giuliani2008quantum}: the up-up component $\langle\rho_\uparrow(\mathbf{r})\rho_\uparrow(\mathbf{r}_0)\rangle$ always vanishes at $\mathbf{r}_0$ forming the \textit{exchange hole} -- see Fig.~\ref{fig:pair}(b). Thus, for a polarized gas the overall PCF should be lower at $\mathbf{r}_0$ than for an unpolarized one. 
For interacting holes the PCF can be greater than 1, see the down-up component $\langle\rho_\downarrow(\mathbf{r})\rho_\uparrow(\mathbf{r}_0)\rangle$ in Fig.~\ref{fig:pair}(b,c). In the noninteracting case, Fig.~\ref{fig:pair}(d), the down-up component is constant. 
What is more interesting for weaker interactions ($\epsilon>5$) is that we observe the characteristic valley antiferromagnetic (AFM) ordering with alternating spin-up/down ($K'$/$K$) densities, see Fig.~\ref{fig:pair}(c), also identified in MoS$_2$ QDs~\cite{Szulakowska_Hawrylak_2020}. 
It is important to note that due to the spin-valley locking effect (where each valley is associated with a specific spin) in TMD materials, spin ordering automatically results in spatial valley ordering for holes.  Consequently, we refer to this ordering as valley-AFM, analogous to spin antiferromagnetism. The typical spatial separation between neighboring valley domains along the channel, as shown in Fig.~\ref{fig:pair}(c), is approximately $\tau_K=30$~nm.

The valley polarization and valley AFM ordering rely on the strong spin-valley locking effect for holes, which is characteristic of most TMDs. Therefore, we expect similar results for other TMDs used as a base material.

In Fig.~\ref{fig:pair}(a) there is also presented an example spin-density $\langle\hat{\rho}_\sigma(\mathbf{r})\rangle$ that respects translational invariance along the channel direction. The CI state for which the spin-density is presented in Fig.~\ref{fig:pair}(a) is partially polarized, $P=2$, with spin-up density $\rho_\uparrow$ of larger amplitude than spin-down $\rho_\downarrow$.

\subsection{Zigzag Wigner crystallization}
\begin{figure}[b]
    \centering
    \includegraphics[width=.48\textwidth]{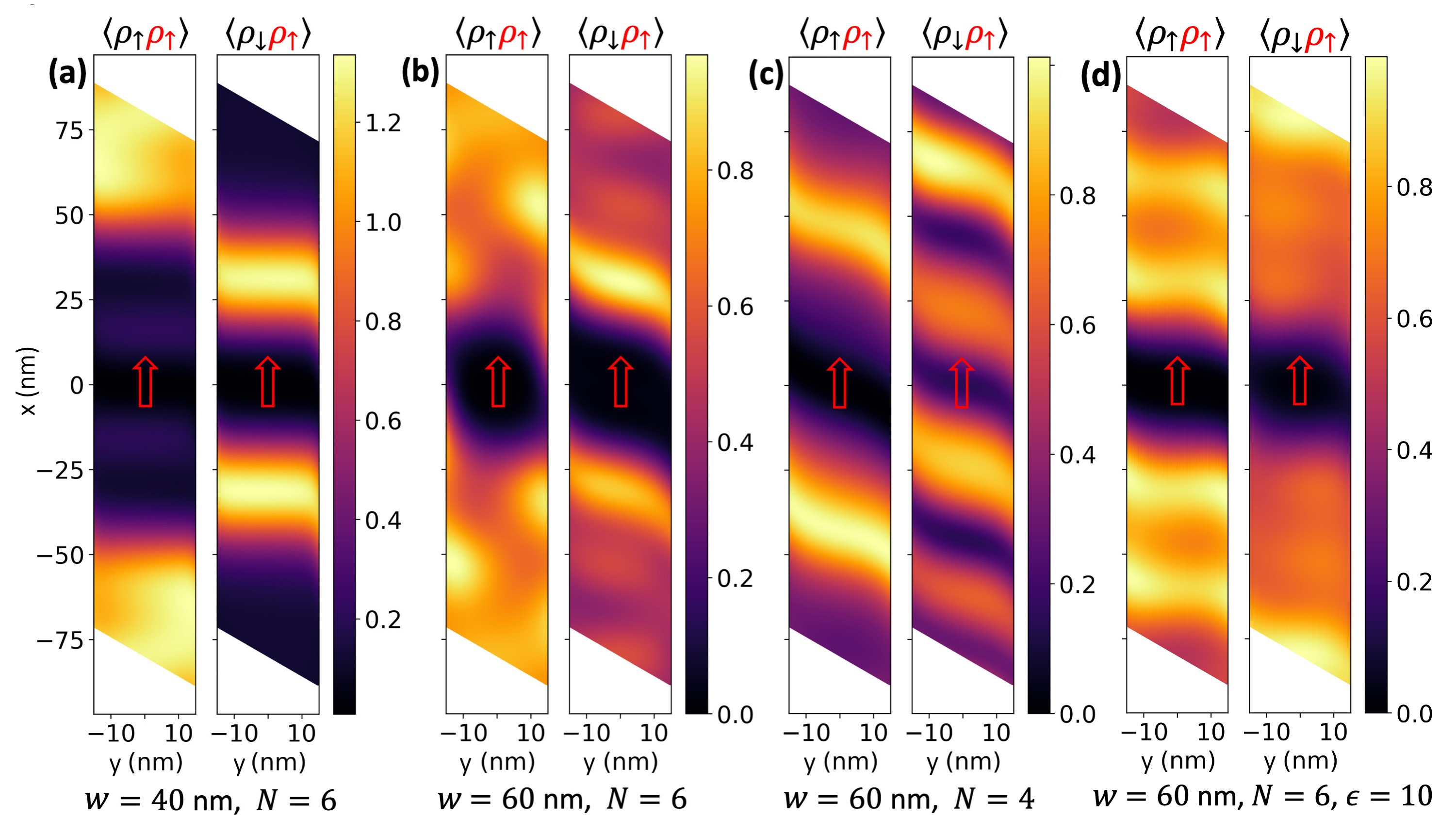}
    \caption{\label{fig:wigner}Spin-spin  pair correlation functions for wider channels: (a) $w=40$~nm, and (b-d) $w=60$~nm. For sufficiently wide channels (b), the Coulomb interaction (here $\epsilon=1$) dominates the kinetic energy and we observe the zigzag Wigner phase. Zigzag ordering disappears for (c) a smaller density of holes ($N=4$), or (d) a weaker Coulomb interaction ($\epsilon=10$).}
\end{figure}

Let us now analyze the effect of reducing the hole density. In 2DEG at low density ($r_s\gtrsim35$), the potential energy dominates the kinetic one and  Wigner crystal  forms~\cite{Attaccalite2002}. In 1D channel it is also possible to observe the Wigner crystallization~\cite{Guclu_Baranger_2009,Guclu2016,Goldberg2024}. However, in a quasi 1D system another ordering emerges at the boundary between these two phases. When an additional energy scale defined by the confinement potential
$U^\mathrm{ch}$
is present, it becomes energetically favorable for holes to move away from the center of the channel when the confinement energy and the Coulomb repulsion are equal: $U^\mathrm{ch}(r_0)=\frac{e^2}{\epsilon r_0}$~\cite{Meyer_2009}. This leads to the \textit{zigzag} Wigner phase~\cite{Piacente2004,Matveev_2004,Klironomos2007,Mehta_Baranger_2013,Goldberg2024} for $r_s\gtrsim r_0$. The zigzag ordering appears for a higher density (lower $r_s$) than the linear ordering. 
Wigner crystallization was recently reported in TMDs-based devices in a junction of two nearby MoSe$_2$ monolayers~\cite{Zhou2021}; and directly observed using STM imaging: 
generalized Wigner crystal in twisted WSe$_2$/WS$_2$ heterostructure~\cite{Hongyuan2021} and twisted WS$_2$ bilayer~\cite{Li2024a} moir\'e superlattice, Wigner molecular crystals in twisted WS$_2$~\cite{li2024b}, and linear 1D Wigner crystal in channels formed at layer-stacking domain walls (under applied strain) in bilayer WS$_2$~\cite{li2024c}.

The characteristic distance $r_s$ for the hole liquid in the channel ($w=30$~nm) is about $\sim185/\epsilon$ (in units of Bohr radius) for the density of $3\times 10^{11}$~cm$^{-2}$ ($N=6$).
To verify whether it is possible to observe the Wigner crystallization in our  device,
we reduced the hole density and analyzed the behavior of the PCF~\cite{Wang2012}.
One should also note that due to the conservation of parity along the transverse direction $y$ (associated with the symmetric confinement $U^\mathrm{ch}(y)$) direct observation of the zigzag ordering is not possible in CI charge density~\cite{Goldberg2024}, therefore, one needs to employ the PCF.

In Fig.~\ref{fig:wigner} we show plots of the spin-resolved PCF for wider channels with $w=40$~nm, Fig.~\ref{fig:wigner}(a), and $w=60$~nm, Fig.~\ref{fig:wigner}(b-d). Decreasing the density twofold, as in Fig.~\ref{fig:wigner}(b), results in the emergence of the zigzag configuration, especially in the $\langle\rho_\uparrow\rho_\uparrow\rangle$ component of PCF. However, further decreasing the density by reducing the number of holes to $N=4$ as in Fig.~\ref{fig:wigner}(c) results rather in a linear configuration instead of the zigzag. Decreasing the interaction strength by setting $\epsilon=10$, as in Fig.~\ref{fig:wigner}(d), results in getting back to the standard hole liquid. 
Both the linear and the zigzag Wigner-crystallization orderings were also revealed in elongated nanostructures using the UHF method~\cite{Goldberg2024}.

Regarding magnetic properties, in 1D crystals typically the antiferromagnetic Heisenberg chain is formed~\cite{Meyer_2009}. On the other hand, the phase diagram of the zigzag spin chain is more complicated and has a number of unpolarized phases as well as regions of complete and partial spin polarization~\cite{Meyer_2009}. Here for zigzag ordering we do not have clear AFM, in Fig.~\ref{fig:wigner}(b) the $\langle\rho_\downarrow\rho_\uparrow\rangle$ component exhibits rather a linear than zigzag ordering present in $\langle\rho_\uparrow\rho_\uparrow\rangle$.
AFM phase is more clearly visible for the linear ordering, see $\langle\rho_\downarrow\rho_\uparrow\rangle$ in Fig.~\ref{fig:wigner}(c).
The additional conductance plateau corresponding to the value of $0.5e^2/h$~\cite{Justin2023} can be connected with the formation of a zigzag structure~\cite{Kumar2019}.
Such a fractional quantized conductance in a quasi-1D electron system was also observed in GaAs/AlGaAs heterojunction~\cite{Kumar2019,Kumar2019A}.

\section{Summary}
In summary, we presented here a theory of interacting valence holes in a gate-defined one dimensional quantum channel in a single layer of transition metal dichalcogenide material WSe$_2$. We utilized a microscopic atomistic tight-binding model to
calculate the hole single-particle energies and wave functions, and employed the Hartree-Fock and configuration-interaction tools to account for hole-hole interactions. For strongly interacting holes, we showed the formation of symmetry-broken, valley-polarized ground states.
We analyzed the interplay of interactions, a perpendicular magnetic field, and a lateral confinement asymmetry together with a strong Rashba spin-orbit coupling present in the WSe$_2$ material and studied their impact on valley polarization.
For strong interactions, our investigation of the pair correlation function revealed the existence of a valley-antiferromagnetic phase. For a lower hole density, we predicted the formation of a zigzag Wigner crystal. We also discussed the impact of various hole liquid phases on transport in a high mobility quasi-one dimensional channel.

\section{Acknowledgements}
JP and MB have been supported by National Science Centre, Poland, under Grant No. 2021/43/D/ST3/01989.
We gratefully acknowledge Polish high-performance computing infrastructure PLGrid (ACK Cyfronet AGH) for providing computer facilities and support within computational grant no. PLG/2022/015633.
DM and PH were supported by NSERC Discovery
Grant No. RGPIN 2019-05714, 
NSERC Alliance Quantum Grant No. ALLRP/578466-2022,
the QSP-078 project of the Quantum Sensors Program at the National Research
Council of Canada, University of Ottawa Research
Chair in Quantum Theory of Materials, Nanostructures,
and Devices, and Digital Research Alliance Canada with computing resources.

\setcounter{figure}{0}
\renewcommand{\thefigure}{S\arabic{figure}}
\appendix
\section*{Appendixes}

\section{Tight-binding parameters}
In the tight-binding single-particle calculations we assume an $11$-band model with even and odd W and Se$_2$ dimer orbitals~\cite{Bieniek_Hawrylak_2018}. We use the following structural parameters: $d_{\parallel}=1.9188$~\AA~(W-Se$_2$ dimer center distance), $d_{\perp}=1.6792$~\AA~(Se atom distance from $z=0$ plane). The tight-binding Slater-Koster parameters are given by (all in eV): $E_d = -0.4330 $, $E_{p0} = -3.8219$, $E_{p\pm 1} = -2.3760$, $V_{dp\sigma}=-1.58193$, $V_{dp\pi}=1.17505$, $V_{dd\sigma}=-0.90501$, $V_{dd\pi}=1.0823$, $V_{dd\delta}=-0.1056$, $V_{pp\sigma}= 0.52091$, $V_{pp\pi}=-0.16775$, together with intrinsic spin-orbit couplings: $\lambda_\mathrm{W}=0.275$, $\lambda_{\mathrm{Se}_2}=0.020$. We note that these parameters are chosen to reproduce the ab initio electronic band structure of two spinful valence bands. 
\begin{figure}[bt]
    \centering
    \includegraphics[width=.45\textwidth]{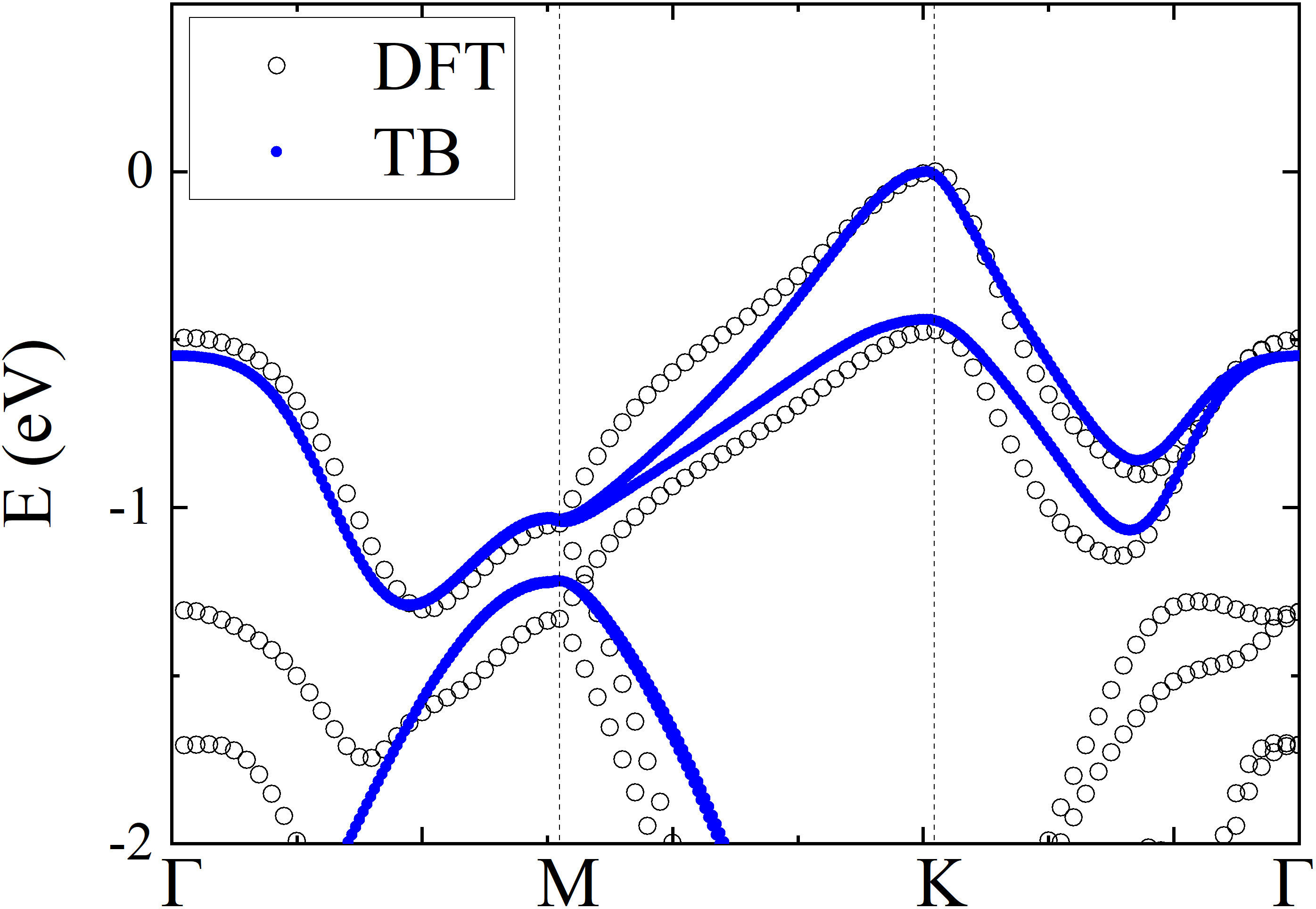}
    \caption{\label{fig:dft_tb} Tight-binding bands (blue) together with reference DFT calculations (dots) showing top of the VB for WSe$_2$ monolayer.
 }
\end{figure}
Fig.~\ref{fig:dft_tb} shows tight-binding~\cite{Bieniek_Hawrylak_2020} bands obtained for these parameters together with reference DFT bands calculated for WSe$_2$ monolayer.
DFT calculations were performed using PBE version of GGA pseudopotential as implemented in Abinit 8.10.2 software~\cite{abinit}. 
Vacuum between layers was set to $10$~\AA, kinetic energy cut-off was taken as 30~Ha, $12\times12\times1$ $k$-point grid, atomic positions were relaxed with stopping criterion of maximal force set to 1$0^{-5}$~Ha/Bohr and total energy difference between self-consistent steps was converged below $10^{-10}$~Ha.

\section{Analytical considerations}
\subsection{Minimal four-state model}
\begin{figure}[bt]
    \centering
    \includegraphics[width=.45\textwidth]{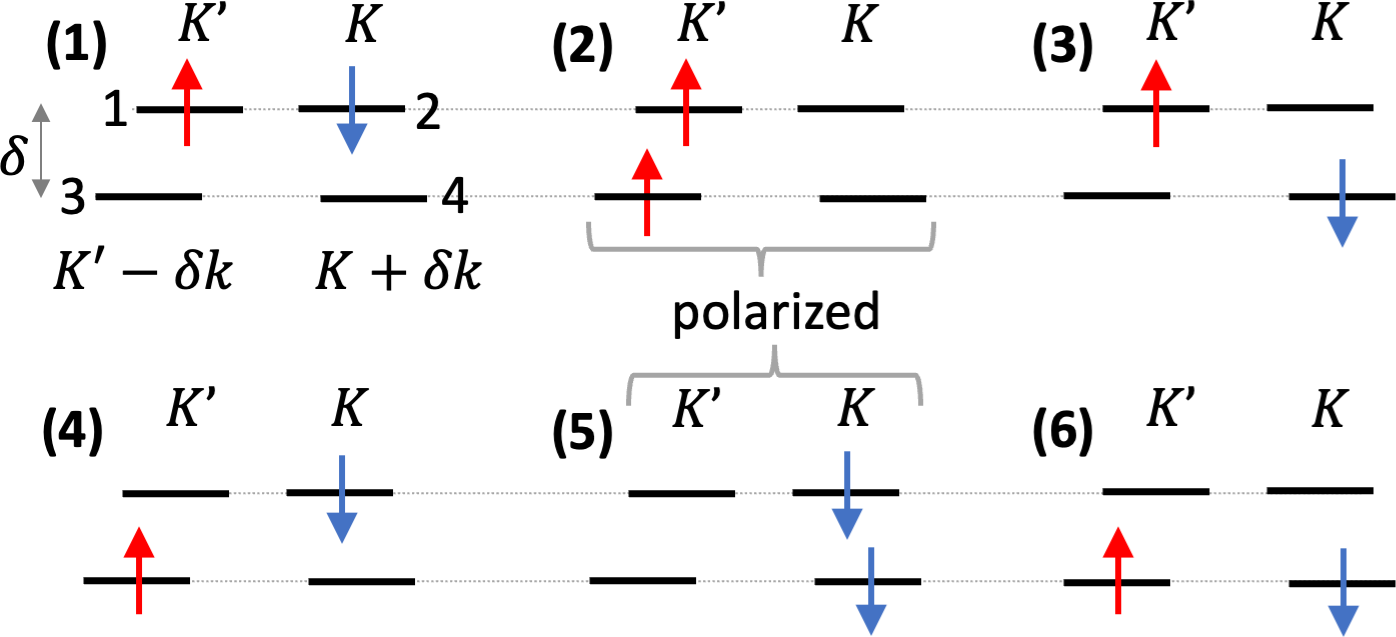}
    \caption{\label{fig:4b_ci}Minimal model with $N=2$ holes occupying 4 states (6 configurations) located on top of both valleys.
 }
\end{figure}

Let us build a minimal analytical Hamiltonian defined in a basis of six configurations (for $N=2$) listed in Fig.~\ref{fig:4b_ci}.
Diagonal elements are composed of direct terms $V_d$ that contribute to the energy with positive signs, and exchange terms $V_x$ with negative sign~\cite{Szabo}, e.g. $\langle{}13|31\rangle$, nonzero only for configurations of the same spins. Among the off-diagonal elements, the only one of importance is the intervalley element $\langle{}12|34\rangle=V_{ir}$.
All other elements can be found in Appendix, Section~C.
Therefore, 
the minimal interacting block can be composed only using configurations (1), (2), and (6) (with the second polarized pair -- (5) omitted for brevity):
\begin{equation}
H_\mathrm{4B}=
\frac{1}{\epsilon}
\begin{pmatrix}
    V_d & 0 & V_{ir}\\
    0 & V_d-V_x+\delta & 0\\
    V_{ir} & 0 & V_d+2\delta
\end{pmatrix}.
\end{equation}
The Hamiltonian $H_\mathrm{4B}$ can be easily diagonalized with two unpolarized states: $|\!\uparrow\downarrow\rangle^\pm$, and one polarized $|\!\uparrow\uparrow\rangle$, with eigenenergies:
\begin{align}\label{eq:4stat_spec}
&E^{\pm}_{\uparrow\downarrow}=\frac{1}{\epsilon}(V_d\pm\sqrt{V_{ir}^2+\delta^2\epsilon^2})+\delta,\\
&E_{\uparrow\uparrow}=\frac{1}{\epsilon}(V_d-V_x)+\delta.
\end{align}
For $\epsilon\rightarrow\infty$, the single unpolarized state
$|\!\uparrow\downarrow\rangle^-$, with the eigenenergy $E^{-}_{\uparrow\downarrow}$, is the GS, well separated in energy from other states. For $V_x=V_{ir}\gg\delta\epsilon$, as in our system, polarized and unpolarized states are almost degenerate.
In case of $V_x>V_{ir}\gg\delta\epsilon$, GS is strictly polarized. 
The relationship between the intravalley exchange $V_x$ and the intervalley scattering $V_{ir}$ elements is crucial and, as it will be shown in the next section, their equality follows from the underlying symmetries present in the system.

\subsection{Conditions for emergence of triplet GS for interacting holes}
In the following section we will demonstrate that, due to symmetries in Coulomb matrix elements originating from the time-reversal symmetry (TRS) in the system, a valley triplet composed of two polarized and one unpolarized states will always appear as the GS in the presence of strong interactions.
The Coulomb matrix elements labeled using channel wavevector number $q_i$ of single-particle components $\Psi^{s,q_i,\sigma}$ splits into two families:
\begin{align}
&V_\mathrm{intra}^{q_1,q_2,q_3,q_4}=\langle k+q_1, k+q_2|V/\epsilon|k+q_3,k+q_4\rangle,\nonumber\\ 
&V_\mathrm{inter}^{q_1,q_2,q_3,q_4}=\langle k+q_1,-k+q_2|V/\epsilon|k+q_3,-k+q_4\rangle,\nonumber  
\end{align}
where $q_i$ is defined as a wavevector shift in the Coulomb integral with respect to valley minima: $k\in\{K,K'\}$.
For every diagonal exchange element (intravalley, because the intervalley exchange vanishes)  $V_\mathrm{intra}^{q,q'\!,q'\!,q}$
which lowers the energy of the polarized configuration, there always exists an off-diagonal (intervalley) scattering element $V_\mathrm{inter}^{q,q,q'\!,q'}$
which couples two intervalley, i.e., unpolarized configurations, with the same amplitude,
\begin{equation}\label{eq:intra_inter}
V_\mathrm{intra}^{q,q'\!,q'\!,q}=V_\mathrm{inter}^{q,q,q'\!,q'}.
\end{equation}
For example, $\langle13|31\rangle=\langle12|34\rangle\simeq56.3$~meV. Now we will show how this is connected with the TRS in the system. The TRS together with the translational symmetry along the channel gives
$B^{\mathrm{VB},s}_{-k_1,k_2,-\sigma}=B^{\ast\,\mathrm{VB},s}_{k_1,k_2,\sigma}$. 
Combining it with a standard TRS of the Bloch bands 
$A^{\mathrm{VB}}_{-\mathbf{k},-\sigma,-l}=A^{\ast\,\mathrm{VB}}_{\mathbf{k},\sigma,l}$~\cite{dresselhaus2007group} 
leads to the symmetry of single-particle states (the Kramers theorem):
\begin{equation}\label{eq:sing_sym}
\Psi^{s,-k,-\sigma}=\Psi^{\ast\,s,k,\sigma}.
\end{equation}
After using the latter symmetry in the Coulomb elements (see Appendix, Section~C) we arrive at a general statement,
\begin{equation}\label{eq:intra_inter_spin}
V^{q,\sigma;q'\!,\sigma'\!;q'\!,\sigma;q,\sigma'}_\mathrm{intra}=V^{q,\sigma;q,-\sigma'\!;q'\!,\sigma;q'\!,-\sigma'}_\mathrm{inter},
\end{equation}
which in our case ($\sigma=\sigma'$) just reduces to Eq.~(\ref{eq:intra_inter}) but with an explicitly written requirement that intervalley elements couple states with opposite spin.
The symmetry in Coulomb matrix elements ensures that every polarized state is accompanied by an unpolarized one, as can be seen in Fig.~\ref{fig:ci}(c).


\section{Coulomb matrix elements}
\begin{figure}[bt]
    \centering
    \includegraphics[width=.4\textwidth]{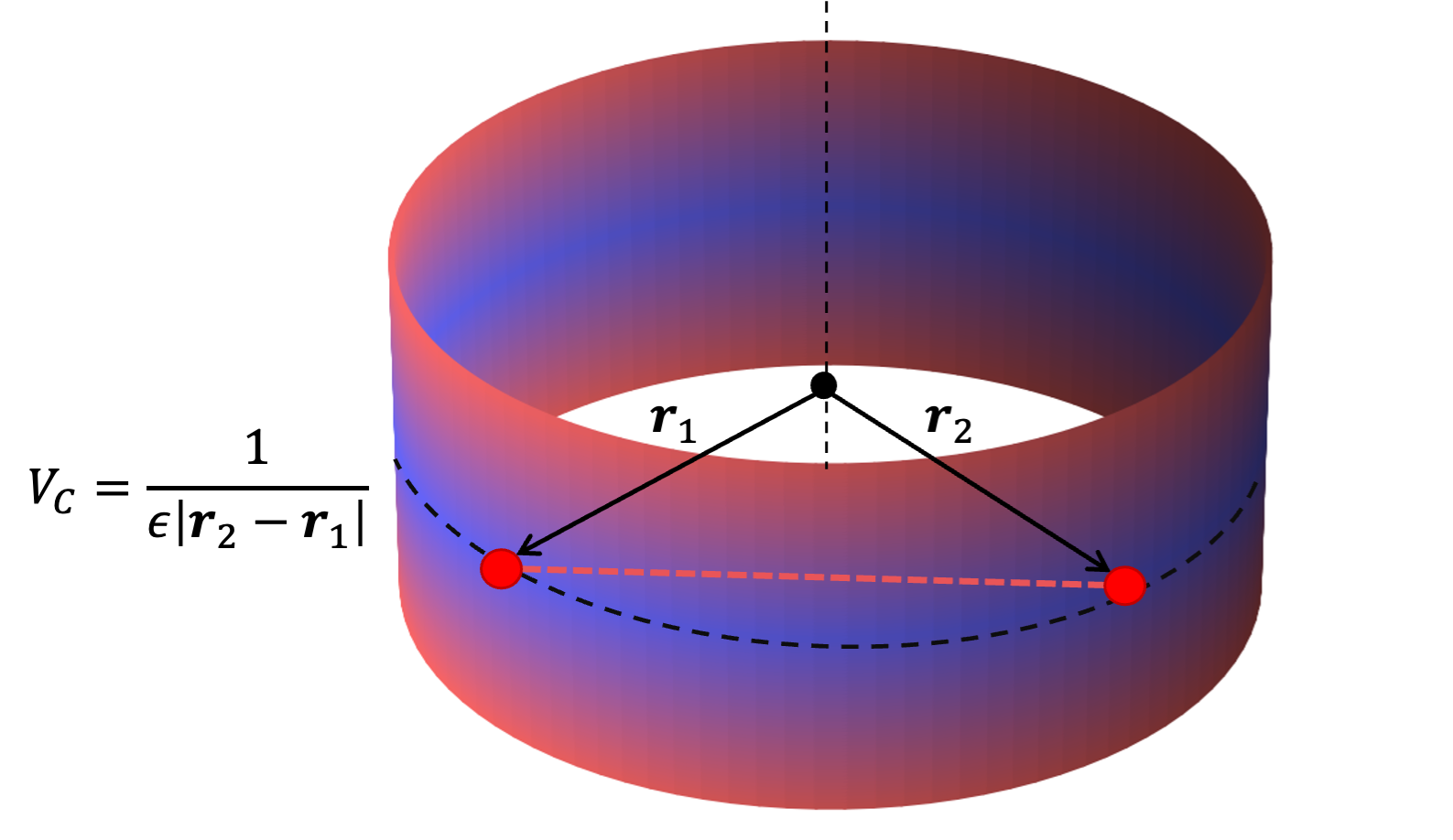}
    \caption{\label{fig:ring} Coulomb interaction (at a distance marked by a red dashed line) on a ring made by rolling up the computational rhombus.  
 }
\end{figure}
In the calculations we adopt a Coulomb interaction model that automatically respects periodic boundary conditions: we roll the computational rhombus into a ring by joining opposite sides -- see Fig.~\ref{fig:ring}. A similar approach was used in~\cite{Mehta_Baranger_2013, Guclu_Baranger_2009}. Now the distance taken in the Coulomb integral is just the section (red dashed line) between two points on the ring. 
The Coulomb matrix elements $V_{ijkl}$ were calculated in exact manner between single-particle channel hole states $\Psi^{s,k_1,\sigma}\equiv\Psi_i$, Eq.~(\ref{eq:singehole}), as expressed by Eq.~(\ref{eq:coulombelement}) (for details see also methodology in [\onlinecite{pawlowski_2021}, Eqs.~(9) and (10)] which we follow here) and then used in the configuration-interaction Hamiltonian Eq.~(\ref{eq:Hci}).
\begin{figure}[b]
    \centering
    \includegraphics[width=.45\textwidth]{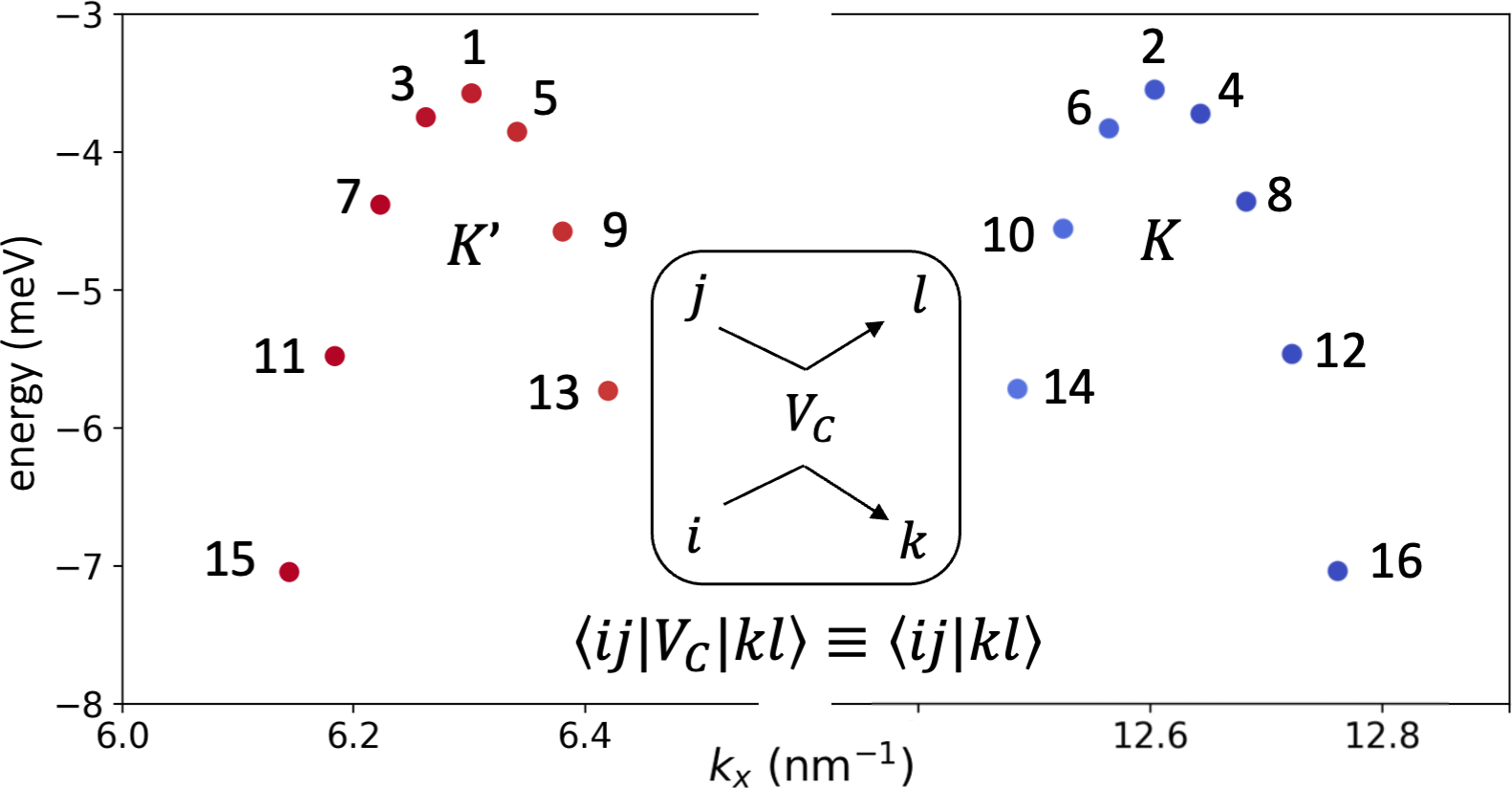}
    \caption{\label{fig:numbering} Numbering scheme in Coulomb matrix elements.  
 }
\end{figure}
When calculating the matrix elements $V_{ijkl}$ we included the leading order one- and two-center integrals~\cite{pawlowski_2021}.
For atomic Coulomb-matrix elements, calculated between atomic nodes, we take exact values for on-site up to next nearest-neighbors exactly (calculated using a Monte-Carlo approach with hydrogen-like Slater orbitals~\cite{Clementi1967}), while the remaining long-range elements are taken as a classical point-like density-density Coulomb interaction between charges on a ring -- see Fig.~\ref{fig:ring}.
The explicit form for the Coulomb integral is:
\begin{align}\label{eq:coulombelement}
&V_{ijkl}\equiv\langle ij|V/\epsilon|kl\rangle=\\
&=\iint d^3\mathbf{r}_1 d^3\mathbf{r}_2
\left(\Psi_i(\mathbf{r}_1)\Psi_j(\mathbf{r}_2)\right)^\dag
\frac{1}{\epsilon|\mathbf{r}_2-\mathbf{r}_1|}
\Psi_k(\mathbf{r}_1)\Psi_l(\mathbf{r}_2)\nonumber\\
&=\iint d^3\mathbf{r}_1 d^3\mathbf{r}_2
\left(\Psi^{s_1,k_1,\sigma}(\mathbf{r}_1)\otimes\Psi^{s_2,k_2,\sigma'}(\mathbf{r}_2)\right)^\dag
\frac{1}{\epsilon|\mathbf{r}_2-\mathbf{r}_1|}\nonumber\\
&\hspace{2.3cm}\times\Psi^{s_3,k_3,\sigma}(\mathbf{r}_1)\otimes\Psi^{s_4,k_4,\sigma'}(\mathbf{r}_2)\nonumber\\
&=\sum_{k'_1,k'_2,k'_3,k'_4} \left(B^{\mathrm{VB},s}_{k_1,k'_1,\sigma}B^{\mathrm{VB},s}_{k_2,k'_2,\sigma'}\right)^\ast B^{\mathrm{VB},s}_{k_3,k'_3,\sigma}B^{\mathrm{VB},s}_{k_4,k'_4,\sigma'}\nonumber\\
&\times\iint d^3\mathbf{r}_1 d^3\mathbf{r}_2\left(\psi^\mathrm{VB}_{k_1,k'_1,\sigma}(\mathbf{r}_1)\otimes\psi^\mathrm{VB}_{k_2,k'_2,\sigma'}(\mathbf{r}_2)\right)^\dag
\frac{1}{\epsilon|\mathbf{r}_2-\mathbf{r}_1|}\nonumber\\
&\hspace{2.2cm}\times\psi^\mathrm{VB}_{k_3,k'_3,\sigma}(\mathbf{r}_1)\otimes\psi^\mathrm{VB}_{k_4,k'_4,\sigma'}(\mathbf{r}_2),\nonumber
\end{align}
where $\mathbf{r}$ is evaluated on a ring as presented in Fig.~\ref{fig:ring}.
The numbering scheme for basis states assumed in the Coulomb elements is presented in Fig.~\ref{fig:numbering}.

Among different matrix elements, those which conserve momentum (elastic scattering), i.e., $q_1+q_2=q_3+q_4$ in $V_\mathrm{intra}^{q_1,q_2,q_3,q_4}$ and $V_\mathrm{inter}^{q_1,q_2,q_3,q_4}$, have the highest amplitudes.
The intravalley direct as well as exchange elements can be nonzero. However, intervalley exchange elements always vanish, for example $\langle 12|21\rangle=0$.
Equality between $V_\mathrm{inter}$ and $V_\mathrm{intra}$ coined in Eq.~(\ref{eq:intra_inter_spin}) can be further generalized on a broader family of integrals:
$V_\mathrm{intra}^{q_1,q_2,q_3,q_4}$ is always accompanied by $V_\mathrm{inter}^{q_1,q_1,q_3,q_3}$ and
$V_\mathrm{inter}^{q_2,q_2,q_4,q_4}$,
and also by
$V_\mathrm{inter}^{q_1,q_4,q_3,q_2}$ and $V_\mathrm{inter}^{q_2,q_3,q_4,q_1}$. 
For example: $\langle{}3,5|7,9\rangle=\langle{}3,4|7,8\rangle=\langle{}5,8|9,4\rangle\simeq 56.3$~meV, and 
$\langle{}5,6|9,10\rangle=\langle{}3,10|7,6\rangle\simeq -56.2$~meV.
We also verified that fundamental symmetries~\cite{Szabo}, i.e., $\langle ij|kl\rangle=\langle ji|lk\rangle$ and $\langle ij|kl\rangle=\langle kl|ij\rangle^\ast$ always holds.

Now let us list the elements considered when building the four-state model in Appendix, Section~B.
The diagonal ones are composed of direct terms: 
intervalley:
$\langle1,2|1,2\rangle\simeq\langle1,4|1,4\rangle\simeq\langle2,3|2,3\rangle\simeq\langle3,4|3,4\rangle\simeq V_d=69.7$~meV, and intravalley: $\langle1,3|1,3\rangle\simeq\langle2,4|2,4\rangle\simeq V_d=69.7$~meV.
The diagonal exchange terms are only intravalley -- with the same spin:
$\langle1,3|3,1\rangle\simeq\langle2,4|4,2\rangle\simeq V_x=56.3$~meV.
Largest off-diagonal (i.e. $>2$~meV), both inter: $\langle1,2|3,4\rangle=V_{ir}\simeq56.3$~meV, $\langle1,4|3,2\rangle=V'_{ir}\simeq4.6$~meV. 
Values for some other Coulomb matrix elements are presented in Table~\ref{tab:coulomb_matrix}.
\begin{table}[]
\begin{tabular}{|lll|}
\hline
\multicolumn{1}{|l|}{integral} & \multicolumn{1}{l|}{value (meV)} & $q$ conserving \\ \hline \hline
\multicolumn{3}{|c|}{\bf between intravalley pairs} \\ \hline
\multicolumn{1}{|l|}{$\langle{}1,1|1,1\rangle$} & \multicolumn{1}{l|}{69.661} & yes \\ \hline
\multicolumn{1}{|l|}{$\langle{}1,3|1,3\rangle$} & \multicolumn{1}{l|}{69.706} & yes \\ \hline
\multicolumn{1}{|l|}{$\langle{}1,3|3,1\rangle$} & \multicolumn{1}{l|}{56.310} & yes \\ \hline
\multicolumn{1}{|l|}{$\langle{}1,1|3,5\rangle$} & \multicolumn{1}{l|}{$-56.269$} & almost \\ \hline
\multicolumn{1}{|l|}{$\langle{}1,3|7,5\rangle$} & \multicolumn{1}{l|}{37.736} & almost \\ \hline
\multicolumn{1}{|l|}{$\langle{}1,1|5,7\rangle$} & \multicolumn{1}{l|}{0.143} & no \\ \hline
\multicolumn{1}{|l|}{$\langle{}1,1|3,7\rangle$} & \multicolumn{1}{l|}{2.967} & no \\ \hline \hline
\multicolumn{3}{|c|}{\bf between intervalley pairs} \\ \hline 
\multicolumn{1}{|l|}{$\langle{}1,2|1,2\rangle$} & \multicolumn{1}{l|}{69.661} & yes \\ \hline
\multicolumn{1}{|l|}{$\langle{}1,2|3,4\rangle$} & \multicolumn{1}{l|}{56.310} & yes \\ \hline
\multicolumn{1}{|l|}{$\langle{}1,4|3,2\rangle$} & \multicolumn{1}{l|}{4.549} & no \\ \hline
\multicolumn{1}{|l|}{$\langle{}1,2|3,6\rangle$} & \multicolumn{1}{l|}{4.548} & no \\ \hline
\multicolumn{1}{|l|}{$\langle{}1,6|3,2\rangle$} & \multicolumn{1}{l|}{$-56.269$} & almost \\ \hline
\multicolumn{1}{|l|}{$\langle{}1,2|11,12\rangle$} & \multicolumn{1}{l|}{$-32.892$} & yes \\ \hline
\multicolumn{1}{|l|}{$\langle{}1,12|11,2\rangle$} & \multicolumn{1}{l|}{1.123} & no \\ \hline
\multicolumn{1}{|l|}{$\langle{}1,14|11,2\rangle$} & \multicolumn{1}{l|}{32.781} & almost \\ \hline
\end{tabular}
\caption{Example Coulomb matrix elements.}
\label{tab:coulomb_matrix}
\end{table}

When calculating the Coulomb integrals we assumed $\epsilon_\mathrm{WSe_2}=5$ and a “quasi-Keldysh” screening: with no screening onsite ($\epsilon=1$), half screening for nearest neighbors and next-nearest neighbors, and a normal screening for others (e.g. $\epsilon=5$). 
The CI basis was composed out of 16 states from the first mode (8 for each valley), as presented in Fig.~\ref{fig:numbering}, plus 8 states from the second mode ($N_b=24$ in total). To verify the calculations convergence: 14+14+2 ($N_b=30$) states were used with very similar results. Moreover, our analysis for variable number of configurations in CI also shows convergence of the results.

\section{Hole density and pair correlation function}
Let us define the hole density $\rho(\mathbf{r})=\langle\Psi|\hat{\rho}(\mathbf{r})|\Psi\rangle$, where the particle density operator: $\hat{\rho}(\mathbf{r})=\sum_{a=1}^N\delta(\mathbf{r}_a-\mathbf{r})$, and $N$ is the number of particles (holes).
Let $|\Psi^{\nu}\rangle$ be expanded in the CI basis of occupation number states (configurations) $p$: $|\Psi^{\nu}\rangle = \sum_{p=1}^M A^{\nu}_p|p\rangle$, and the number of configurations $M=\binom{N_b}{N}$, where $N_b$ is the number of single-particle states taken to construct the $N$-particle subspace of the Fock space.
Any single-particle $\sum_{a=1}^N O_1(\mathbf{r}_a)$ and two-particle $\frac{1}{2}\sum_{a\neq b} O_2(\mathbf{r}_a,\mathbf{r}_b)$ operator in the second quantization representation reads:
\begin{equation}
\begin{split}
&\hat{\mathcal{O}}_1=\sum_{i,j}\langle i|O_1|j\rangle h^\dag_i h_j,\\
&\hat{\mathcal{O}}_2=\frac{1}{2}\sum_{i,j,k,l}\langle ij|O_2|kl\rangle h^\dag_i h^\dag_j h_l h_k.
\end{split}
\end{equation}
The density operator $\hat{\rho}(\mathbf{r})$ depends on a single coordinate only, thus it is a single-particle operator.
Now we can express the hole density $\rho(\mathbf{r})$ using our CI basis within the second quantization representation (we don't have to worry about proper antisymmetry of basis states -- the Fock states are implicitly antisymmetric):
\begin{equation}
\begin{split}
\rho(\mathbf{r})=\langle\Psi^{\nu}|\hat{\rho}(\mathbf{r})|\Psi^{\nu}\rangle=
\sum_{p,q}A_p^{\nu\ast} A^{\nu}_q \sum_{i,j} \langle p|h^\dag_i h_j|q\rangle
\langle i|\hat{\rho}(\mathbf{r})|j\rangle,
\end{split}
\end{equation}
where the matrix element $\langle i|\hat{\rho}(\mathbf{r}) |j\rangle=\int d^2\mathbf{r}_1\Psi^\dag_i(\mathbf{r}_1)\delta(\mathbf{r}_1-\mathbf{r})\Psi_j(\mathbf{r}_1)=\Psi^\dag_i(\mathbf{r})\Psi_j(\mathbf{r})$.
To evaluate the elements $\langle p| h^\dag_i h_j|q\rangle$  we have to do simple algebra for a string of creation/annihilation operators or equivalently use the Slater-Condon rules~\cite{Szabo}.
Accordingly, the two-particle pair correlation operator $\frac{1}{2}\!\sum_{a\neq b}\delta(\mathbf{r}_a-\mathbf{r})\delta(\mathbf{r}_b-\mathbf{r}_0)$ can be defined as:
\begin{align}
\langle\hat{\rho}&(\mathbf{r})\hat{\rho}(\mathbf{r}_0)\rangle=\\
&=\frac{1}{4}\sum_{p,q}A^{\nu\ast}_p A^{\nu}_q \sum_{i,j,k,l} \langle p| h^\dag_i h^\dag_j h_l h_k|q\rangle
\langle ij|\hat{\rho}(\mathbf{r})\hat{\rho}(\mathbf{r}_0)|kl\rangle,\nonumber
\end{align}
where 
\begin{align}
&\langle ij|\hat{\rho}(\mathbf{r})\hat{\rho}(\mathbf{r}_0)|kl\rangle=\\
&=\Psi^\dag_i(\mathbf{r})\Psi^\dag_j(\mathbf{r}_0)\Psi_k(\mathbf{r})\Psi_l(\mathbf{r}_0)+\Psi^\dag_i(\mathbf{r}_0)\Psi^\dag_j(\mathbf{r})\Psi_k(\mathbf{r}_0)\Psi_l(\mathbf{r}).\nonumber
\end{align}
We can also take only the given spin coordinates: 
\begin{align}
\langle ij|\hat{\rho}_\sigma&(\mathbf{r})\hat{\rho}_{\sigma'}(\mathbf{r}_0)|kl\rangle=\\
&=\Psi^\dag_{i\sigma}(\mathbf{r})\Psi^\dag_{j\sigma'}(\mathbf{r}_0)\Psi_{k\sigma}(\mathbf{r})\Psi_{l\sigma'}(\mathbf{r}_0)+\nonumber\\
&+\Psi^\dag_{i\sigma'}(\mathbf{r}_0)\Psi^\dag_{j\sigma}(\mathbf{r})\Psi_{k\sigma'}(\mathbf{r}_0)\Psi_{l\sigma}(\mathbf{r}),\nonumber
\end{align}
and apply a normalization factor $\frac{1}{\langle \hat{\rho}(\mathbf{r})\rangle \langle \hat{\rho}(\mathbf{r}_0)\rangle}$. The normalization ensures that
$\langle\rho_\uparrow(\mathbf{r})\rho_\uparrow(\mathbf{r}_0)\rangle$
and 
$\langle\rho_\downarrow(\mathbf{r})\rho_\uparrow(\mathbf{r}_0)\rangle$
in a limit of large distances, $|\mathbf{r}-\mathbf{r}_0|\rightarrow\infty$, approach 1.

\section{Momentum of correlated states}

\begin{figure}[tb]
    \vspace{3mm}
    \centering
    \includegraphics[width=.35\textwidth]{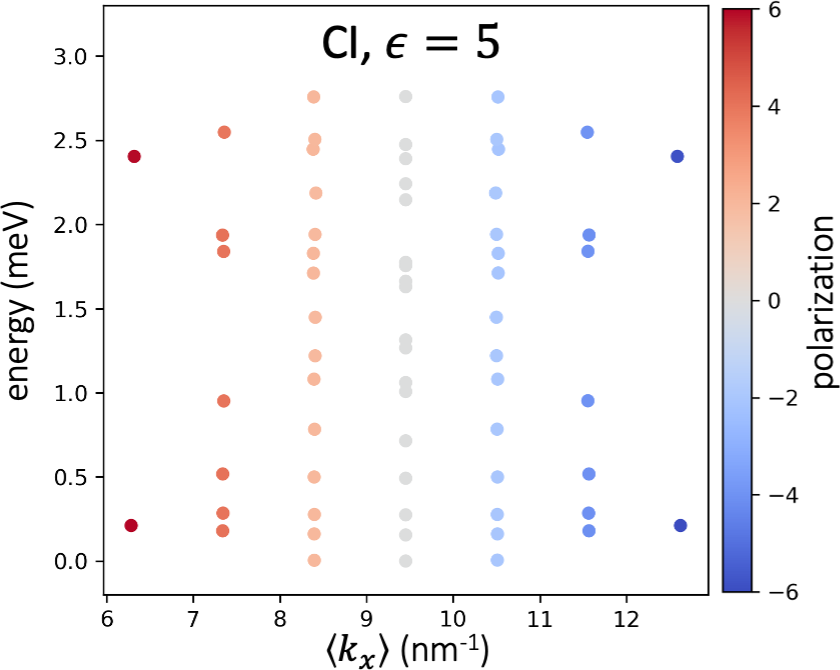}
    \caption{\label{fig:ci_k} Energies of (first 66) correlated states of $N=6$ holes for $\epsilon=5$ plotted against their $k_x$ expectation value.
 }
\end{figure}

Finally, in Fig.~\ref{fig:ci_k} we present a ladder of many-body hole states plotted against their expectation value of momentum $\langle{}k_x\rangle$ along the channel.
Maximally polarized states are localized exactly in their respective valleys, while all other states are located in between.
Here, again, it is clearly visible that due to intra- and intervalley elements equilibrium, the polarized states are always accompanied by unpolarized ones.

\bibliography{main}

\begin{thebibliography}{71}%
\makeatletter
\providecommand \@ifxundefined [1]{%
 \@ifx{#1\undefined}
}%
\providecommand \@ifnum [1]{%
 \ifnum #1\expandafter \@firstoftwo
 \else \expandafter \@secondoftwo
 \fi
}%
\providecommand \@ifx [1]{%
 \ifx #1\expandafter \@firstoftwo
 \else \expandafter \@secondoftwo
 \fi
}%
\providecommand \natexlab [1]{#1}%
\providecommand \enquote  [1]{``#1''}%
\providecommand \bibnamefont  [1]{#1}%
\providecommand \bibfnamefont [1]{#1}%
\providecommand \citenamefont [1]{#1}%
\providecommand \href@noop [0]{\@secondoftwo}%
\providecommand \href [0]{\begingroup \@sanitize@url \@href}%
\providecommand \@href[1]{\@@startlink{#1}\@@href}%
\providecommand \@@href[1]{\endgroup#1\@@endlink}%
\providecommand \@sanitize@url [0]{\catcode `\\12\catcode `\$12\catcode
  `\&12\catcode `\#12\catcode `\^12\catcode `\_12\catcode `\%12\relax}%
\providecommand \@@startlink[1]{}%
\providecommand \@@endlink[0]{}%
\providecommand \url  [0]{\begingroup\@sanitize@url \@url }%
\providecommand \@url [1]{\endgroup\@href {#1}{\urlprefix }}%
\providecommand \urlprefix  [0]{URL }%
\providecommand \Eprint [0]{\href }%
\providecommand \doibase [0]{https://doi.org/}%
\providecommand \selectlanguage [0]{\@gobble}%
\providecommand \bibinfo  [0]{\@secondoftwo}%
\providecommand \bibfield  [0]{\@secondoftwo}%
\providecommand \translation [1]{[#1]}%
\providecommand \BibitemOpen [0]{}%
\providecommand \bibitemStop [0]{}%
\providecommand \bibitemNoStop [0]{.\EOS\space}%
\providecommand \EOS [0]{\spacefactor3000\relax}%
\providecommand \BibitemShut  [1]{\csname bibitem#1\endcsname}%
\let\auto@bib@innerbib\@empty
\bibitem [{\citenamefont {Goh}\ \emph {et~al.}(2020)\citenamefont {Goh},
  \citenamefont {Bussolotti}, \citenamefont {Lau}, \citenamefont
  {Kotekar-Patil}, \citenamefont {Ooi},\ and\ \citenamefont {Chee}}]{Goh2019}%
  \BibitemOpen
  \bibfield  {author} {\bibinfo {author} {\bibfnamefont {K.~E.~J.}\
  \bibnamefont {Goh}}, \bibinfo {author} {\bibfnamefont {F.}~\bibnamefont
  {Bussolotti}}, \bibinfo {author} {\bibfnamefont {C.~S.}\ \bibnamefont {Lau}},
  \bibinfo {author} {\bibfnamefont {D.}~\bibnamefont {Kotekar-Patil}}, \bibinfo
  {author} {\bibfnamefont {Z.~E.}\ \bibnamefont {Ooi}},\ and\ \bibinfo {author}
  {\bibfnamefont {J.~Y.}\ \bibnamefont {Chee}},\ }\bibfield  {title} {\bibinfo
  {title} {Toward valley-coupled spin qubits},\ }\href
  {https://doi.org/https://doi.org/10.1002/qute.201900123} {\bibfield
  {journal} {\bibinfo  {journal} {Advanced Quantum Technologies}\ }\textbf
  {\bibinfo {volume} {3}},\ \bibinfo {pages} {1900123} (\bibinfo {year}
  {2020})}\BibitemShut {NoStop}%
\bibitem [{\citenamefont {Sierra}\ \emph {et~al.}(2021)\citenamefont {Sierra},
  \citenamefont {Fabian}, \citenamefont {Kawakami}, \citenamefont {Roche},\
  and\ \citenamefont {Valenzuela}}]{Sierra2021}%
  \BibitemOpen
  \bibfield  {author} {\bibinfo {author} {\bibfnamefont {J.~F.}\ \bibnamefont
  {Sierra}}, \bibinfo {author} {\bibfnamefont {J.}~\bibnamefont {Fabian}},
  \bibinfo {author} {\bibfnamefont {R.~K.}\ \bibnamefont {Kawakami}}, \bibinfo
  {author} {\bibfnamefont {S.}~\bibnamefont {Roche}},\ and\ \bibinfo {author}
  {\bibfnamefont {S.~O.}\ \bibnamefont {Valenzuela}},\ }\bibfield  {title}
  {\bibinfo {title} {Van der {Waals} heterostructures for spintronics and
  opto-spintronics},\ }\href {https://doi.org/10.1038/s41565-021-00936-x}
  {\bibfield  {journal} {\bibinfo  {journal} {Nature Nanotechnology}\ }\textbf
  {\bibinfo {volume} {16}},\ \bibinfo {pages} {856} (\bibinfo {year}
  {2021})}\BibitemShut {NoStop}%
\bibitem [{\citenamefont {Liu}\ and\ \citenamefont {Hersam}(2019)}]{Liu2019}%
  \BibitemOpen
  \bibfield  {author} {\bibinfo {author} {\bibfnamefont {X.}~\bibnamefont
  {Liu}}\ and\ \bibinfo {author} {\bibfnamefont {M.~C.}\ \bibnamefont
  {Hersam}},\ }\bibfield  {title} {\bibinfo {title} {{2D} materials for quantum
  information science},\ }\href {https://doi.org/10.1038/s41578-019-0136-x}
  {\bibfield  {journal} {\bibinfo  {journal} {Nature Reviews Materials}\
  }\textbf {\bibinfo {volume} {4}},\ \bibinfo {pages} {669} (\bibinfo {year}
  {2019})}\BibitemShut {NoStop}%
\bibitem [{\citenamefont {Xiao}\ \emph {et~al.}(2012)\citenamefont {Xiao},
  \citenamefont {Liu}, \citenamefont {Feng}, \citenamefont {Xu},\ and\
  \citenamefont {Yao}}]{Xiao2012}%
  \BibitemOpen
  \bibfield  {author} {\bibinfo {author} {\bibfnamefont {D.}~\bibnamefont
  {Xiao}}, \bibinfo {author} {\bibfnamefont {G.-B.}\ \bibnamefont {Liu}},
  \bibinfo {author} {\bibfnamefont {W.}~\bibnamefont {Feng}}, \bibinfo {author}
  {\bibfnamefont {X.}~\bibnamefont {Xu}},\ and\ \bibinfo {author}
  {\bibfnamefont {W.}~\bibnamefont {Yao}},\ }\bibfield  {title} {\bibinfo
  {title} {Coupled spin and valley physics in monolayers of {M}o{S}$_{2}$ and
  other group-vi dichalcogenides},\ }\href
  {https://doi.org/10.1103/PhysRevLett.108.196802} {\bibfield  {journal}
  {\bibinfo  {journal} {Phys. Rev. Lett.}\ }\textbf {\bibinfo {volume} {108}},\
  \bibinfo {pages} {196802} (\bibinfo {year} {2012})}\BibitemShut {NoStop}%
\bibitem [{\citenamefont {Gong}\ \emph {et~al.}(2013)\citenamefont {Gong},
  \citenamefont {Liu}, \citenamefont {Yu}, \citenamefont {Xiao}, \citenamefont
  {Cui}, \citenamefont {Xu},\ and\ \citenamefont {Yao}}]{Gong2013}%
  \BibitemOpen
  \bibfield  {author} {\bibinfo {author} {\bibfnamefont {Z.}~\bibnamefont
  {Gong}}, \bibinfo {author} {\bibfnamefont {G.-B.}\ \bibnamefont {Liu}},
  \bibinfo {author} {\bibfnamefont {H.}~\bibnamefont {Yu}}, \bibinfo {author}
  {\bibfnamefont {D.}~\bibnamefont {Xiao}}, \bibinfo {author} {\bibfnamefont
  {X.}~\bibnamefont {Cui}}, \bibinfo {author} {\bibfnamefont {X.}~\bibnamefont
  {Xu}},\ and\ \bibinfo {author} {\bibfnamefont {W.}~\bibnamefont {Yao}},\
  }\bibfield  {title} {\bibinfo {title} {Magnetoelectric effects and
  valley-controlled spin quantum gates in transition metal dichalcogenide
  bilayers},\ }\href {https://doi.org/10.1038/ncomms3053} {\bibfield  {journal}
  {\bibinfo  {journal} {Nature Communications}\ }\textbf {\bibinfo {volume}
  {4}},\ \bibinfo {pages} {2053} (\bibinfo {year} {2013})}\BibitemShut
  {NoStop}%
\bibitem [{\citenamefont {Volmer}\ \emph {et~al.}(2023)\citenamefont {Volmer},
  \citenamefont {Ersfeld}, \citenamefont {Faria~Junior}, \citenamefont
  {Waldecker}, \citenamefont {Parashar}, \citenamefont {Rathmann},
  \citenamefont {Dubey}, \citenamefont {Cojocariu}, \citenamefont {Feyer},
  \citenamefont {Watanabe}, \citenamefont {Taniguchi}, \citenamefont
  {Schneider}, \citenamefont {Plucinski}, \citenamefont {Stampfer},
  \citenamefont {Fabian},\ and\ \citenamefont {Beschoten}}]{Volmer2023}%
  \BibitemOpen
  \bibfield  {author} {\bibinfo {author} {\bibfnamefont {F.}~\bibnamefont
  {Volmer}}, \bibinfo {author} {\bibfnamefont {M.}~\bibnamefont {Ersfeld}},
  \bibinfo {author} {\bibfnamefont {P.~E.}\ \bibnamefont {Faria~Junior}},
  \bibinfo {author} {\bibfnamefont {L.}~\bibnamefont {Waldecker}}, \bibinfo
  {author} {\bibfnamefont {B.}~\bibnamefont {Parashar}}, \bibinfo {author}
  {\bibfnamefont {L.}~\bibnamefont {Rathmann}}, \bibinfo {author}
  {\bibfnamefont {S.}~\bibnamefont {Dubey}}, \bibinfo {author} {\bibfnamefont
  {I.}~\bibnamefont {Cojocariu}}, \bibinfo {author} {\bibfnamefont
  {V.}~\bibnamefont {Feyer}}, \bibinfo {author} {\bibfnamefont
  {K.}~\bibnamefont {Watanabe}}, \bibinfo {author} {\bibfnamefont
  {T.}~\bibnamefont {Taniguchi}}, \bibinfo {author} {\bibfnamefont {C.~M.}\
  \bibnamefont {Schneider}}, \bibinfo {author} {\bibfnamefont {L.}~\bibnamefont
  {Plucinski}}, \bibinfo {author} {\bibfnamefont {C.}~\bibnamefont {Stampfer}},
  \bibinfo {author} {\bibfnamefont {J.}~\bibnamefont {Fabian}},\ and\ \bibinfo
  {author} {\bibfnamefont {B.}~\bibnamefont {Beschoten}},\ }\bibfield  {title}
  {\bibinfo {title} {Twist angle dependent interlayer transfer of valley
  polarization from excitons to free charge carriers in {WSe}2/{MoSe}2
  heterobilayers},\ }\href {https://doi.org/10.1038/s41699-023-00420-1}
  {\bibfield  {journal} {\bibinfo  {journal} {npj 2D Materials and
  Applications}\ }\textbf {\bibinfo {volume} {7}},\ \bibinfo {pages} {58}
  (\bibinfo {year} {2023})}\BibitemShut {NoStop}%
\bibitem [{\citenamefont {Pawłowski}(2019)}]{pawlowski_2019}%
  \BibitemOpen
  \bibfield  {author} {\bibinfo {author} {\bibfnamefont {J.}~\bibnamefont
  {Pawłowski}},\ }\bibfield  {title} {\bibinfo {title} {Spin-valley system in
  a gated {MoS}$_2$-monolayer quantum dot},\ }\href
  {https://doi.org/10.1088/1367-2630/ab5ac9} {\bibfield  {journal} {\bibinfo
  {journal} {New Journal of Physics}\ }\textbf {\bibinfo {volume} {21}},\
  \bibinfo {pages} {123029} (\bibinfo {year} {2019})}\BibitemShut {NoStop}%
\bibitem [{\citenamefont {Alt\ifmmode \imath \else \i
  \fi{}nta\ifmmode~\mbox{\c{s}}\else \c{s}\fi{}}\ \emph
  {et~al.}(2021)\citenamefont {Alt\ifmmode \imath \else \i
  \fi{}nta\ifmmode~\mbox{\c{s}}\else \c{s}\fi{}}, \citenamefont {Bieniek},
  \citenamefont {Dusko}, \citenamefont {Korkusi\ifmmode~\acute{n}\else
  \'{n}\fi{}ski}, \citenamefont {Paw\l{}owski},\ and\ \citenamefont
  {Hawrylak}}]{Altintas_Hawrylak_2021}%
  \BibitemOpen
  \bibfield  {author} {\bibinfo {author} {\bibfnamefont {A.}~\bibnamefont
  {Alt\ifmmode \imath \else \i \fi{}nta\ifmmode~\mbox{\c{s}}\else \c{s}\fi{}}},
  \bibinfo {author} {\bibfnamefont {M.}~\bibnamefont {Bieniek}}, \bibinfo
  {author} {\bibfnamefont {A.}~\bibnamefont {Dusko}}, \bibinfo {author}
  {\bibfnamefont {M.}~\bibnamefont {Korkusi\ifmmode~\acute{n}\else
  \'{n}\fi{}ski}}, \bibinfo {author} {\bibfnamefont {J.}~\bibnamefont
  {Paw\l{}owski}},\ and\ \bibinfo {author} {\bibfnamefont {P.}~\bibnamefont
  {Hawrylak}},\ }\bibfield  {title} {\bibinfo {title} {Spin-valley qubits in
  gated quantum dots in a single layer of transition metal dichalcogenides},\
  }\href {https://doi.org/10.1103/PhysRevB.104.195412} {\bibfield  {journal}
  {\bibinfo  {journal} {Phys. Rev. B}\ }\textbf {\bibinfo {volume} {104}},\
  \bibinfo {pages} {195412} (\bibinfo {year} {2021})}\BibitemShut {NoStop}%
\bibitem [{\citenamefont {Paw\l{}owski}\ \emph {et~al.}(2021)\citenamefont
  {Paw\l{}owski}, \citenamefont {Bieniek},\ and\ \citenamefont
  {Wo\ifmmode~\acute{z}\else \'{z}\fi{}niak}}]{pawlowski_2021}%
  \BibitemOpen
  \bibfield  {author} {\bibinfo {author} {\bibfnamefont {J.}~\bibnamefont
  {Paw\l{}owski}}, \bibinfo {author} {\bibfnamefont {M.}~\bibnamefont
  {Bieniek}},\ and\ \bibinfo {author} {\bibfnamefont {T.}~\bibnamefont
  {Wo\ifmmode~\acute{z}\else \'{z}\fi{}niak}},\ }\bibfield  {title} {\bibinfo
  {title} {Valley two-qubit system in a {MoS}$_{2}$-monolayer gated double
  quantum dot},\ }\href {https://doi.org/10.1103/PhysRevApplied.15.054025}
  {\bibfield  {journal} {\bibinfo  {journal} {Phys. Rev. Applied}\ }\textbf
  {\bibinfo {volume} {15}},\ \bibinfo {pages} {054025} (\bibinfo {year}
  {2021})}\BibitemShut {NoStop}%
\bibitem [{\citenamefont {Rohling}\ and\ \citenamefont
  {Burkard}(2012)}]{Rohling_2012}%
  \BibitemOpen
  \bibfield  {author} {\bibinfo {author} {\bibfnamefont {N.}~\bibnamefont
  {Rohling}}\ and\ \bibinfo {author} {\bibfnamefont {G.}~\bibnamefont
  {Burkard}},\ }\bibfield  {title} {\bibinfo {title} {Universal quantum
  computing with spin and valley states},\ }\href
  {https://doi.org/10.1088/1367-2630/14/8/083008} {\bibfield  {journal}
  {\bibinfo  {journal} {New Journal of Physics}\ }\textbf {\bibinfo {volume}
  {14}},\ \bibinfo {pages} {083008} (\bibinfo {year} {2012})}\BibitemShut
  {NoStop}%
\bibitem [{\citenamefont {Burkard}\ \emph {et~al.}(2023)\citenamefont
  {Burkard}, \citenamefont {Ladd}, \citenamefont {Pan}, \citenamefont
  {Nichol},\ and\ \citenamefont {Petta}}]{Burkard2023}%
  \BibitemOpen
  \bibfield  {author} {\bibinfo {author} {\bibfnamefont {G.}~\bibnamefont
  {Burkard}}, \bibinfo {author} {\bibfnamefont {T.~D.}\ \bibnamefont {Ladd}},
  \bibinfo {author} {\bibfnamefont {A.}~\bibnamefont {Pan}}, \bibinfo {author}
  {\bibfnamefont {J.~M.}\ \bibnamefont {Nichol}},\ and\ \bibinfo {author}
  {\bibfnamefont {J.~R.}\ \bibnamefont {Petta}},\ }\bibfield  {title} {\bibinfo
  {title} {Semiconductor spin qubits},\ }\href
  {https://doi.org/10.1103/RevModPhys.95.025003} {\bibfield  {journal}
  {\bibinfo  {journal} {Rev. Mod. Phys.}\ }\textbf {\bibinfo {volume} {95}},\
  \bibinfo {pages} {025003} (\bibinfo {year} {2023})}\BibitemShut {NoStop}%
\bibitem [{\citenamefont {Ciorga}\ \emph {et~al.}(2000)\citenamefont {Ciorga},
  \citenamefont {Sachrajda}, \citenamefont {Hawrylak}, \citenamefont {Gould},
  \citenamefont {Zawadzki}, \citenamefont {Jullian}, \citenamefont {Feng},\
  and\ \citenamefont {Wasilewski}}]{Ciorga2000}%
  \BibitemOpen
  \bibfield  {author} {\bibinfo {author} {\bibfnamefont {M.}~\bibnamefont
  {Ciorga}}, \bibinfo {author} {\bibfnamefont {A.~S.}\ \bibnamefont
  {Sachrajda}}, \bibinfo {author} {\bibfnamefont {P.}~\bibnamefont {Hawrylak}},
  \bibinfo {author} {\bibfnamefont {C.}~\bibnamefont {Gould}}, \bibinfo
  {author} {\bibfnamefont {P.}~\bibnamefont {Zawadzki}}, \bibinfo {author}
  {\bibfnamefont {S.}~\bibnamefont {Jullian}}, \bibinfo {author} {\bibfnamefont
  {Y.}~\bibnamefont {Feng}},\ and\ \bibinfo {author} {\bibfnamefont
  {Z.}~\bibnamefont {Wasilewski}},\ }\bibfield  {title} {\bibinfo {title}
  {Addition spectrum of a lateral dot from coulomb and spin-blockade
  spectroscopy},\ }\href {https://doi.org/10.1103/PhysRevB.61.R16315}
  {\bibfield  {journal} {\bibinfo  {journal} {Phys. Rev. B}\ }\textbf {\bibinfo
  {volume} {61}},\ \bibinfo {pages} {R16315} (\bibinfo {year}
  {2000})}\BibitemShut {NoStop}%
\bibitem [{\citenamefont {Allen}\ \emph {et~al.}(2012)\citenamefont {Allen},
  \citenamefont {Martin},\ and\ \citenamefont {Yacoby}}]{Allen2012}%
  \BibitemOpen
  \bibfield  {author} {\bibinfo {author} {\bibfnamefont {M.~T.}\ \bibnamefont
  {Allen}}, \bibinfo {author} {\bibfnamefont {J.}~\bibnamefont {Martin}},\ and\
  \bibinfo {author} {\bibfnamefont {A.}~\bibnamefont {Yacoby}},\ }\bibfield
  {title} {\bibinfo {title} {Gate-defined quantum confinement in suspended
  bilayer graphene},\ }\href {https://doi.org/10.1038/ncomms1945} {\bibfield
  {journal} {\bibinfo  {journal} {Nature Communications}\ }\textbf {\bibinfo
  {volume} {3}},\ \bibinfo {pages} {934} (\bibinfo {year} {2012})}\BibitemShut
  {NoStop}%
\bibitem [{\citenamefont {Eich}\ \emph {et~al.}(2018)\citenamefont {Eich},
  \citenamefont {Herman}, \citenamefont {Pisoni}, \citenamefont {Overweg},
  \citenamefont {Kurzmann}, \citenamefont {Lee}, \citenamefont {Rickhaus},
  \citenamefont {Watanabe}, \citenamefont {Taniguchi}, \citenamefont {Sigrist},
  \citenamefont {Ihn},\ and\ \citenamefont {Ensslin}}]{Eich2018}%
  \BibitemOpen
  \bibfield  {author} {\bibinfo {author} {\bibfnamefont {M.}~\bibnamefont
  {Eich}}, \bibinfo {author} {\bibfnamefont {F.}~\bibnamefont {Herman}},
  \bibinfo {author} {\bibfnamefont {R.}~\bibnamefont {Pisoni}}, \bibinfo
  {author} {\bibfnamefont {H.}~\bibnamefont {Overweg}}, \bibinfo {author}
  {\bibfnamefont {A.}~\bibnamefont {Kurzmann}}, \bibinfo {author}
  {\bibfnamefont {Y.}~\bibnamefont {Lee}}, \bibinfo {author} {\bibfnamefont
  {P.}~\bibnamefont {Rickhaus}}, \bibinfo {author} {\bibfnamefont
  {K.}~\bibnamefont {Watanabe}}, \bibinfo {author} {\bibfnamefont
  {T.}~\bibnamefont {Taniguchi}}, \bibinfo {author} {\bibfnamefont
  {M.}~\bibnamefont {Sigrist}}, \bibinfo {author} {\bibfnamefont
  {T.}~\bibnamefont {Ihn}},\ and\ \bibinfo {author} {\bibfnamefont
  {K.}~\bibnamefont {Ensslin}},\ }\bibfield  {title} {\bibinfo {title} {Spin
  and valley states in gate-defined bilayer graphene quantum dots},\ }\href
  {https://doi.org/10.1103/PhysRevX.8.031023} {\bibfield  {journal} {\bibinfo
  {journal} {Phys. Rev. X}\ }\textbf {\bibinfo {volume} {8}},\ \bibinfo {pages}
  {031023} (\bibinfo {year} {2018})}\BibitemShut {NoStop}%
\bibitem [{\citenamefont {Banszerus}\ \emph {et~al.}(2020)\citenamefont
  {Banszerus}, \citenamefont {M{\"o}ller}, \citenamefont {Icking},
  \citenamefont {Watanabe}, \citenamefont {Taniguchi}, \citenamefont {Volk},\
  and\ \citenamefont {Stampfer}}]{Banszerus2020}%
  \BibitemOpen
  \bibfield  {author} {\bibinfo {author} {\bibfnamefont {L.}~\bibnamefont
  {Banszerus}}, \bibinfo {author} {\bibfnamefont {S.}~\bibnamefont
  {M{\"o}ller}}, \bibinfo {author} {\bibfnamefont {E.}~\bibnamefont {Icking}},
  \bibinfo {author} {\bibfnamefont {K.}~\bibnamefont {Watanabe}}, \bibinfo
  {author} {\bibfnamefont {T.}~\bibnamefont {Taniguchi}}, \bibinfo {author}
  {\bibfnamefont {C.}~\bibnamefont {Volk}},\ and\ \bibinfo {author}
  {\bibfnamefont {C.}~\bibnamefont {Stampfer}},\ }\bibfield  {title} {\bibinfo
  {title} {Single-electron double quantum dots in bilayer graphene},\ }\href
  {https://doi.org/10.1021/acs.nanolett.9b05295} {\bibfield  {journal}
  {\bibinfo  {journal} {Nano Letters}\ }\textbf {\bibinfo {volume} {20}},\
  \bibinfo {pages} {2005} (\bibinfo {year} {2020})}\BibitemShut {NoStop}%
\bibitem [{\citenamefont {Tong}\ \emph {et~al.}(2021)\citenamefont {Tong},
  \citenamefont {Garreis}, \citenamefont {Knothe}, \citenamefont {Eich},
  \citenamefont {Sacchi}, \citenamefont {Watanabe}, \citenamefont {Taniguchi},
  \citenamefont {Fal'ko}, \citenamefont {Ihn}, \citenamefont {Ensslin},\ and\
  \citenamefont {Kurzmann}}]{Tong2021}%
  \BibitemOpen
  \bibfield  {author} {\bibinfo {author} {\bibfnamefont {C.}~\bibnamefont
  {Tong}}, \bibinfo {author} {\bibfnamefont {R.}~\bibnamefont {Garreis}},
  \bibinfo {author} {\bibfnamefont {A.}~\bibnamefont {Knothe}}, \bibinfo
  {author} {\bibfnamefont {M.}~\bibnamefont {Eich}}, \bibinfo {author}
  {\bibfnamefont {A.}~\bibnamefont {Sacchi}}, \bibinfo {author} {\bibfnamefont
  {K.}~\bibnamefont {Watanabe}}, \bibinfo {author} {\bibfnamefont
  {T.}~\bibnamefont {Taniguchi}}, \bibinfo {author} {\bibfnamefont
  {V.}~\bibnamefont {Fal'ko}}, \bibinfo {author} {\bibfnamefont
  {T.}~\bibnamefont {Ihn}}, \bibinfo {author} {\bibfnamefont {K.}~\bibnamefont
  {Ensslin}},\ and\ \bibinfo {author} {\bibfnamefont {A.}~\bibnamefont
  {Kurzmann}},\ }\bibfield  {title} {\bibinfo {title} {Tunable valley splitting
  and bipolar operation in graphene quantum dots},\ }\href
  {https://doi.org/10.1021/acs.nanolett.0c04343} {\bibfield  {journal}
  {\bibinfo  {journal} {Nano Letters}\ }\textbf {\bibinfo {volume} {21}},\
  \bibinfo {pages} {1068} (\bibinfo {year} {2021})}\BibitemShut {NoStop}%
\bibitem [{\citenamefont {Song}\ \emph {et~al.}(2015)\citenamefont {Song},
  \citenamefont {Liu}, \citenamefont {Mosallanejad}, \citenamefont {You},
  \citenamefont {Han}, \citenamefont {Chen}, \citenamefont {Li}, \citenamefont
  {Cao}, \citenamefont {Xiao}, \citenamefont {Guo},\ and\ \citenamefont
  {Guo}}]{Song2015}%
  \BibitemOpen
  \bibfield  {author} {\bibinfo {author} {\bibfnamefont {X.-X.}\ \bibnamefont
  {Song}}, \bibinfo {author} {\bibfnamefont {D.}~\bibnamefont {Liu}}, \bibinfo
  {author} {\bibfnamefont {V.}~\bibnamefont {Mosallanejad}}, \bibinfo {author}
  {\bibfnamefont {J.}~\bibnamefont {You}}, \bibinfo {author} {\bibfnamefont
  {T.-Y.}\ \bibnamefont {Han}}, \bibinfo {author} {\bibfnamefont {D.-T.}\
  \bibnamefont {Chen}}, \bibinfo {author} {\bibfnamefont {H.-O.}\ \bibnamefont
  {Li}}, \bibinfo {author} {\bibfnamefont {G.}~\bibnamefont {Cao}}, \bibinfo
  {author} {\bibfnamefont {M.}~\bibnamefont {Xiao}}, \bibinfo {author}
  {\bibfnamefont {G.-C.}\ \bibnamefont {Guo}},\ and\ \bibinfo {author}
  {\bibfnamefont {G.-P.}\ \bibnamefont {Guo}},\ }\bibfield  {title} {\bibinfo
  {title} {A gate defined quantum dot on the two-dimensional transition metal
  dichalcogenide semiconductor {WSe}$_2$},\ }\href
  {https://doi.org/10.1039/C5NR04961J} {\bibfield  {journal} {\bibinfo
  {journal} {Nanoscale}\ }\textbf {\bibinfo {volume} {7}},\ \bibinfo {pages}
  {16867} (\bibinfo {year} {2015})}\BibitemShut {NoStop}%
\bibitem [{\citenamefont {Pisoni}\ \emph {et~al.}(2018)\citenamefont {Pisoni},
  \citenamefont {Lei}, \citenamefont {Back}, \citenamefont {Eich},
  \citenamefont {Overweg}, \citenamefont {Lee}, \citenamefont {Watanabe},
  \citenamefont {Taniguchi}, \citenamefont {Ihn},\ and\ \citenamefont
  {Ensslin}}]{Pisoni2018}%
  \BibitemOpen
  \bibfield  {author} {\bibinfo {author} {\bibfnamefont {R.}~\bibnamefont
  {Pisoni}}, \bibinfo {author} {\bibfnamefont {Z.}~\bibnamefont {Lei}},
  \bibinfo {author} {\bibfnamefont {P.}~\bibnamefont {Back}}, \bibinfo {author}
  {\bibfnamefont {M.}~\bibnamefont {Eich}}, \bibinfo {author} {\bibfnamefont
  {H.}~\bibnamefont {Overweg}}, \bibinfo {author} {\bibfnamefont
  {Y.}~\bibnamefont {Lee}}, \bibinfo {author} {\bibfnamefont {K.}~\bibnamefont
  {Watanabe}}, \bibinfo {author} {\bibfnamefont {T.}~\bibnamefont {Taniguchi}},
  \bibinfo {author} {\bibfnamefont {T.}~\bibnamefont {Ihn}},\ and\ \bibinfo
  {author} {\bibfnamefont {K.}~\bibnamefont {Ensslin}},\ }\bibfield  {title}
  {\bibinfo {title} {Gate-tunable quantum dot in a high quality single layer
  {MoS}$_2$ {van der Waals} heterostructure},\ }\href
  {https://doi.org/10.1063/1.5021113} {\bibfield  {journal} {\bibinfo
  {journal} {Applied Physics Letters}\ }\textbf {\bibinfo {volume} {112}},\
  \bibinfo {pages} {123101} (\bibinfo {year} {2018})}\BibitemShut {NoStop}%
\bibitem [{\citenamefont {Davari}\ \emph {et~al.}(2020)\citenamefont {Davari},
  \citenamefont {Stacy}, \citenamefont {Mercado}, \citenamefont {Tull},
  \citenamefont {Basnet}, \citenamefont {Pandey}, \citenamefont {Watanabe},
  \citenamefont {Taniguchi}, \citenamefont {Hu},\ and\ \citenamefont
  {Churchill}}]{Davari2020}%
  \BibitemOpen
  \bibfield  {author} {\bibinfo {author} {\bibfnamefont {S.}~\bibnamefont
  {Davari}}, \bibinfo {author} {\bibfnamefont {J.}~\bibnamefont {Stacy}},
  \bibinfo {author} {\bibfnamefont {A.}~\bibnamefont {Mercado}}, \bibinfo
  {author} {\bibfnamefont {J.}~\bibnamefont {Tull}}, \bibinfo {author}
  {\bibfnamefont {R.}~\bibnamefont {Basnet}}, \bibinfo {author} {\bibfnamefont
  {K.}~\bibnamefont {Pandey}}, \bibinfo {author} {\bibfnamefont
  {K.}~\bibnamefont {Watanabe}}, \bibinfo {author} {\bibfnamefont
  {T.}~\bibnamefont {Taniguchi}}, \bibinfo {author} {\bibfnamefont
  {J.}~\bibnamefont {Hu}},\ and\ \bibinfo {author} {\bibfnamefont
  {H.}~\bibnamefont {Churchill}},\ }\bibfield  {title} {\bibinfo {title}
  {Gate-defined accumulation-mode quantum dots in monolayer and bilayer
  {WSe}$_{2}$},\ }\href {https://doi.org/10.1103/PhysRevApplied.13.054058}
  {\bibfield  {journal} {\bibinfo  {journal} {Phys. Rev. Applied}\ }\textbf
  {\bibinfo {volume} {13}},\ \bibinfo {pages} {054058} (\bibinfo {year}
  {2020})}\BibitemShut {NoStop}%
\bibitem [{\citenamefont {Boddison-Chouinard}\ \emph
  {et~al.}(2021)\citenamefont {Boddison-Chouinard}, \citenamefont {Bogan},
  \citenamefont {Fong}, \citenamefont {Watanabe}, \citenamefont {Taniguchi},
  \citenamefont {Studenikin}, \citenamefont {Sachrajda}, \citenamefont
  {Korkusinski}, \citenamefont {Altintas}, \citenamefont {Bieniek},
  \citenamefont {Hawrylak}, \citenamefont {Luican-Mayer},\ and\ \citenamefont
  {Gaudreau}}]{Chouinard2021}%
  \BibitemOpen
  \bibfield  {author} {\bibinfo {author} {\bibfnamefont {J.}~\bibnamefont
  {Boddison-Chouinard}}, \bibinfo {author} {\bibfnamefont {A.}~\bibnamefont
  {Bogan}}, \bibinfo {author} {\bibfnamefont {N.}~\bibnamefont {Fong}},
  \bibinfo {author} {\bibfnamefont {K.}~\bibnamefont {Watanabe}}, \bibinfo
  {author} {\bibfnamefont {T.}~\bibnamefont {Taniguchi}}, \bibinfo {author}
  {\bibfnamefont {S.}~\bibnamefont {Studenikin}}, \bibinfo {author}
  {\bibfnamefont {A.}~\bibnamefont {Sachrajda}}, \bibinfo {author}
  {\bibfnamefont {M.}~\bibnamefont {Korkusinski}}, \bibinfo {author}
  {\bibfnamefont {A.}~\bibnamefont {Altintas}}, \bibinfo {author}
  {\bibfnamefont {M.}~\bibnamefont {Bieniek}}, \bibinfo {author} {\bibfnamefont
  {P.}~\bibnamefont {Hawrylak}}, \bibinfo {author} {\bibfnamefont
  {A.}~\bibnamefont {Luican-Mayer}},\ and\ \bibinfo {author} {\bibfnamefont
  {L.}~\bibnamefont {Gaudreau}},\ }\bibfield  {title} {\bibinfo {title}
  {{Gate-controlled quantum dots in monolayer {WSe}$_2$}},\ }\href
  {https://doi.org/10.1063/5.0062838} {\bibfield  {journal} {\bibinfo
  {journal} {Applied Physics Letters}\ }\textbf {\bibinfo {volume} {119}},\
  \bibinfo {pages} {133104} (\bibinfo {year} {2021})}\BibitemShut {NoStop}%
\bibitem [{\citenamefont {Kumar}\ \emph {et~al.}(2023)\citenamefont {Kumar},
  \citenamefont {Kim}, \citenamefont {Tripathy}, \citenamefont {Watanabe},
  \citenamefont {Taniguchi}, \citenamefont {Novoselov},\ and\ \citenamefont
  {Kotekar-Patil}}]{Kumar2023}%
  \BibitemOpen
  \bibfield  {author} {\bibinfo {author} {\bibfnamefont {P.}~\bibnamefont
  {Kumar}}, \bibinfo {author} {\bibfnamefont {H.}~\bibnamefont {Kim}}, \bibinfo
  {author} {\bibfnamefont {S.}~\bibnamefont {Tripathy}}, \bibinfo {author}
  {\bibfnamefont {K.}~\bibnamefont {Watanabe}}, \bibinfo {author}
  {\bibfnamefont {T.}~\bibnamefont {Taniguchi}}, \bibinfo {author}
  {\bibfnamefont {K.~S.}\ \bibnamefont {Novoselov}},\ and\ \bibinfo {author}
  {\bibfnamefont {D.}~\bibnamefont {Kotekar-Patil}},\ }\bibfield  {title}
  {\bibinfo {title} {Excited state spectroscopy and spin splitting in single
  layer {MoS}$_2$ quantum dots},\ }\href {https://doi.org/10.1039/D3NR03844K}
  {\bibfield  {journal} {\bibinfo  {journal} {Nanoscale}\ }\textbf {\bibinfo
  {volume} {15}},\ \bibinfo {pages} {18203} (\bibinfo {year}
  {2023})}\BibitemShut {NoStop}%
\bibitem [{\citenamefont {Krishnan}\ \emph {et~al.}(2023)\citenamefont
  {Krishnan}, \citenamefont {Biswas}, \citenamefont {Hsueh}, \citenamefont
  {Ma}, \citenamefont {Rahman},\ and\ \citenamefont {Weber}}]{Krishnan2023}%
  \BibitemOpen
  \bibfield  {author} {\bibinfo {author} {\bibfnamefont {R.}~\bibnamefont
  {Krishnan}}, \bibinfo {author} {\bibfnamefont {S.}~\bibnamefont {Biswas}},
  \bibinfo {author} {\bibfnamefont {Y.-L.}\ \bibnamefont {Hsueh}}, \bibinfo
  {author} {\bibfnamefont {H.}~\bibnamefont {Ma}}, \bibinfo {author}
  {\bibfnamefont {R.}~\bibnamefont {Rahman}},\ and\ \bibinfo {author}
  {\bibfnamefont {B.}~\bibnamefont {Weber}},\ }\bibfield  {title} {\bibinfo
  {title} {Spin-valley locking for in-gap quantum dots in a {MoS}$_2$
  transistor},\ }\href {https://doi.org/10.1021/acs.nanolett.3c01779}
  {\bibfield  {journal} {\bibinfo  {journal} {Nano Letters}\ }\textbf {\bibinfo
  {volume} {23}},\ \bibinfo {pages} {6171–6177} (\bibinfo {year}
  {2023})}\BibitemShut {NoStop}%
\bibitem [{\citenamefont {G\"u\ifmmode~\mbox{\c{c}}\else \c{c}\fi{}l\"u}\ \emph
  {et~al.}(2009{\natexlab{a}})\citenamefont {G\"u\ifmmode~\mbox{\c{c}}\else
  \c{c}\fi{}l\"u}, \citenamefont {Potasz}, \citenamefont {Voznyy},
  \citenamefont {Korkusinski},\ and\ \citenamefont {Hawrylak}}]{Potasz2009}%
  \BibitemOpen
  \bibfield  {author} {\bibinfo {author} {\bibfnamefont {A.~D.}\ \bibnamefont
  {G\"u\ifmmode~\mbox{\c{c}}\else \c{c}\fi{}l\"u}}, \bibinfo {author}
  {\bibfnamefont {P.}~\bibnamefont {Potasz}}, \bibinfo {author} {\bibfnamefont
  {O.}~\bibnamefont {Voznyy}}, \bibinfo {author} {\bibfnamefont
  {M.}~\bibnamefont {Korkusinski}},\ and\ \bibinfo {author} {\bibfnamefont
  {P.}~\bibnamefont {Hawrylak}},\ }\bibfield  {title} {\bibinfo {title}
  {Magnetism and correlations in fractionally filled degenerate shells of
  graphene quantum dots},\ }\href
  {https://doi.org/10.1103/PhysRevLett.103.246805} {\bibfield  {journal}
  {\bibinfo  {journal} {Phys. Rev. Lett.}\ }\textbf {\bibinfo {volume} {103}},\
  \bibinfo {pages} {246805} (\bibinfo {year} {2009}{\natexlab{a}})}\BibitemShut
  {NoStop}%
\bibitem [{\citenamefont {Potasz}\ \emph {et~al.}(2012)\citenamefont {Potasz},
  \citenamefont {G\"u\ifmmode~\mbox{\c{c}}\else \c{c}\fi{}l\"u}, \citenamefont
  {W\'ojs},\ and\ \citenamefont {Hawrylak}}]{Potasz2012}%
  \BibitemOpen
  \bibfield  {author} {\bibinfo {author} {\bibfnamefont {P.}~\bibnamefont
  {Potasz}}, \bibinfo {author} {\bibfnamefont {A.~D.}\ \bibnamefont
  {G\"u\ifmmode~\mbox{\c{c}}\else \c{c}\fi{}l\"u}}, \bibinfo {author}
  {\bibfnamefont {A.}~\bibnamefont {W\'ojs}},\ and\ \bibinfo {author}
  {\bibfnamefont {P.}~\bibnamefont {Hawrylak}},\ }\bibfield  {title} {\bibinfo
  {title} {Electronic properties of gated triangular graphene quantum dots:
  Magnetism, correlations, and geometrical effects},\ }\href
  {https://doi.org/10.1103/PhysRevB.85.075431} {\bibfield  {journal} {\bibinfo
  {journal} {Phys. Rev. B}\ }\textbf {\bibinfo {volume} {85}},\ \bibinfo
  {pages} {075431} (\bibinfo {year} {2012})}\BibitemShut {NoStop}%
\bibitem [{\citenamefont {Shi}\ \emph {et~al.}(2020)\citenamefont {Shi},
  \citenamefont {Shih}, \citenamefont {Gustafsson}, \citenamefont {Rhodes},
  \citenamefont {Kim}, \citenamefont {Watanabe}, \citenamefont {Taniguchi},
  \citenamefont {Papić}, \citenamefont {Hone},\ and\ \citenamefont
  {Dean}}]{Shi2020}%
  \BibitemOpen
  \bibfield  {author} {\bibinfo {author} {\bibfnamefont {Q.}~\bibnamefont
  {Shi}}, \bibinfo {author} {\bibfnamefont {E.-M.}\ \bibnamefont {Shih}},
  \bibinfo {author} {\bibfnamefont {M.~V.}\ \bibnamefont {Gustafsson}},
  \bibinfo {author} {\bibfnamefont {D.~A.}\ \bibnamefont {Rhodes}}, \bibinfo
  {author} {\bibfnamefont {B.}~\bibnamefont {Kim}}, \bibinfo {author}
  {\bibfnamefont {K.}~\bibnamefont {Watanabe}}, \bibinfo {author}
  {\bibfnamefont {T.}~\bibnamefont {Taniguchi}}, \bibinfo {author}
  {\bibfnamefont {Z.}~\bibnamefont {Papić}}, \bibinfo {author} {\bibfnamefont
  {J.}~\bibnamefont {Hone}},\ and\ \bibinfo {author} {\bibfnamefont {C.~R.}\
  \bibnamefont {Dean}},\ }\bibfield  {title} {\bibinfo {title} {Odd- and
  even-denominator fractional quantum hall states in monolayer {WSe}$_2$},\
  }\href {https://doi.org/10.1038/s41565-020-0685-6} {\bibfield  {journal}
  {\bibinfo  {journal} {Nature Nanotechnology}\ }\textbf {\bibinfo {volume}
  {15}},\ \bibinfo {pages} {569–573} (\bibinfo {year} {2020})}\BibitemShut
  {NoStop}%
\bibitem [{\citenamefont {Reddy}\ \emph {et~al.}(2023)\citenamefont {Reddy},
  \citenamefont {Alsallom}, \citenamefont {Zhang}, \citenamefont {Devakul},\
  and\ \citenamefont {Fu}}]{Reddy2023}%
  \BibitemOpen
  \bibfield  {author} {\bibinfo {author} {\bibfnamefont {A.~P.}\ \bibnamefont
  {Reddy}}, \bibinfo {author} {\bibfnamefont {F.}~\bibnamefont {Alsallom}},
  \bibinfo {author} {\bibfnamefont {Y.}~\bibnamefont {Zhang}}, \bibinfo
  {author} {\bibfnamefont {T.}~\bibnamefont {Devakul}},\ and\ \bibinfo {author}
  {\bibfnamefont {L.}~\bibnamefont {Fu}},\ }\bibfield  {title} {\bibinfo
  {title} {Fractional quantum anomalous hall states in twisted bilayer
  {MoTe}$_{2}$ and {WSe}$_{2}$},\ }\href
  {https://doi.org/10.1103/PhysRevB.108.085117} {\bibfield  {journal} {\bibinfo
   {journal} {Phys. Rev. B}\ }\textbf {\bibinfo {volume} {108}},\ \bibinfo
  {pages} {085117} (\bibinfo {year} {2023})}\BibitemShut {NoStop}%
\bibitem [{\citenamefont {Zhao}\ \emph {et~al.}(2023)\citenamefont {Zhao},
  \citenamefont {Huang}, \citenamefont {Crépel}, \citenamefont {Wu},
  \citenamefont {Zhang}, \citenamefont {Wang}, \citenamefont {Han},
  \citenamefont {Li}, \citenamefont {Xi}, \citenamefont {Pan}, \citenamefont
  {Wang}, \citenamefont {Watanabe}, \citenamefont {Taniguchi}, \citenamefont
  {Sacépé}, \citenamefont {Zhang}, \citenamefont {Wang}, \citenamefont {Lu},
  \citenamefont {Regnault},\ and\ \citenamefont {Han}}]{zhao2023}%
  \BibitemOpen
  \bibfield  {author} {\bibinfo {author} {\bibfnamefont {S.}~\bibnamefont
  {Zhao}}, \bibinfo {author} {\bibfnamefont {J.}~\bibnamefont {Huang}},
  \bibinfo {author} {\bibfnamefont {V.}~\bibnamefont {Crépel}}, \bibinfo
  {author} {\bibfnamefont {X.}~\bibnamefont {Wu}}, \bibinfo {author}
  {\bibfnamefont {T.}~\bibnamefont {Zhang}}, \bibinfo {author} {\bibfnamefont
  {H.}~\bibnamefont {Wang}}, \bibinfo {author} {\bibfnamefont {X.}~\bibnamefont
  {Han}}, \bibinfo {author} {\bibfnamefont {Z.}~\bibnamefont {Li}}, \bibinfo
  {author} {\bibfnamefont {C.}~\bibnamefont {Xi}}, \bibinfo {author}
  {\bibfnamefont {S.}~\bibnamefont {Pan}}, \bibinfo {author} {\bibfnamefont
  {Z.}~\bibnamefont {Wang}}, \bibinfo {author} {\bibfnamefont {K.}~\bibnamefont
  {Watanabe}}, \bibinfo {author} {\bibfnamefont {T.}~\bibnamefont {Taniguchi}},
  \bibinfo {author} {\bibfnamefont {B.}~\bibnamefont {Sacépé}}, \bibinfo
  {author} {\bibfnamefont {J.}~\bibnamefont {Zhang}}, \bibinfo {author}
  {\bibfnamefont {N.}~\bibnamefont {Wang}}, \bibinfo {author} {\bibfnamefont
  {J.}~\bibnamefont {Lu}}, \bibinfo {author} {\bibfnamefont {N.}~\bibnamefont
  {Regnault}},\ and\ \bibinfo {author} {\bibfnamefont {Z.~V.}\ \bibnamefont
  {Han}},\ }\href@noop {} {\bibinfo {title} {Probing the fractional quantum
  hall phases in valley-layer locked bilayer {MoS}$_{2}$}} (\bibinfo {year}
  {2023}),\ \Eprint {https://arxiv.org/abs/2308.02821} {arXiv:2308.02821
  [cond-mat.mes-hall]} \BibitemShut {NoStop}%
\bibitem [{\citenamefont {Wang}\ \emph {et~al.}(2017)\citenamefont {Wang},
  \citenamefont {Shan},\ and\ \citenamefont {Mak}}]{Wang2017}%
  \BibitemOpen
  \bibfield  {author} {\bibinfo {author} {\bibfnamefont {Z.}~\bibnamefont
  {Wang}}, \bibinfo {author} {\bibfnamefont {J.}~\bibnamefont {Shan}},\ and\
  \bibinfo {author} {\bibfnamefont {K.~F.}\ \bibnamefont {Mak}},\ }\bibfield
  {title} {\bibinfo {title} {Valley- and spin-polarized {Landau} levels in
  monolayer {WSe}$_2$},\ }\href {https://doi.org/10.1038/nnano.2016.213}
  {\bibfield  {journal} {\bibinfo  {journal} {Nature Nanotechnology}\ }\textbf
  {\bibinfo {volume} {12}},\ \bibinfo {pages} {144} (\bibinfo {year}
  {2017})}\BibitemShut {NoStop}%
\bibitem [{\citenamefont {Scrace}\ \emph {et~al.}(2015)\citenamefont {Scrace},
  \citenamefont {Tsai}, \citenamefont {Barman}, \citenamefont {Schweidenback},
  \citenamefont {Petrou}, \citenamefont {Kioseoglou}, \citenamefont {Ozfidan},
  \citenamefont {Korkusi\'nski},\ and\ \citenamefont
  {Hawrylak}}]{Scrace_Hawrylak_2015}%
  \BibitemOpen
  \bibfield  {author} {\bibinfo {author} {\bibfnamefont {T.}~\bibnamefont
  {Scrace}}, \bibinfo {author} {\bibfnamefont {Y.}~\bibnamefont {Tsai}},
  \bibinfo {author} {\bibfnamefont {B.}~\bibnamefont {Barman}}, \bibinfo
  {author} {\bibfnamefont {L.}~\bibnamefont {Schweidenback}}, \bibinfo {author}
  {\bibfnamefont {A.}~\bibnamefont {Petrou}}, \bibinfo {author} {\bibfnamefont
  {G.}~\bibnamefont {Kioseoglou}}, \bibinfo {author} {\bibfnamefont
  {I.}~\bibnamefont {Ozfidan}}, \bibinfo {author} {\bibfnamefont
  {M.}~\bibnamefont {Korkusi\'nski}},\ and\ \bibinfo {author} {\bibfnamefont
  {P.}~\bibnamefont {Hawrylak}},\ }\bibfield  {title} {\bibinfo {title}
  {{Magnetoluminescence and valley polarized state of a two-dimensional
  electron gas in {WS}$_2$ monolayers}},\ }\href
  {https://doi.org/10.1038/nnano.2015.78} {\bibfield  {journal} {\bibinfo
  {journal} {Nat. Nano.}\ }\textbf {\bibinfo {volume} {10}},\ \bibinfo {pages}
  {603} (\bibinfo {year} {2015})}\BibitemShut {NoStop}%
\bibitem [{\citenamefont {Szulakowska}\ \emph {et~al.}(2020)\citenamefont
  {Szulakowska}, \citenamefont {Cygorek}, \citenamefont {Bieniek},\ and\
  \citenamefont {Hawrylak}}]{Szulakowska_Hawrylak_2020}%
  \BibitemOpen
  \bibfield  {author} {\bibinfo {author} {\bibfnamefont {L.}~\bibnamefont
  {Szulakowska}}, \bibinfo {author} {\bibfnamefont {M.}~\bibnamefont
  {Cygorek}}, \bibinfo {author} {\bibfnamefont {M.}~\bibnamefont {Bieniek}},\
  and\ \bibinfo {author} {\bibfnamefont {P.}~\bibnamefont {Hawrylak}},\
  }\bibfield  {title} {\bibinfo {title} {Valley- and spin-polarized
  broken-symmetry states of interacting electrons in gated {MoS}$_{2}$ quantum
  dots},\ }\href {https://doi.org/10.1103/PhysRevB.102.245410} {\bibfield
  {journal} {\bibinfo  {journal} {Phys. Rev. B}\ }\textbf {\bibinfo {volume}
  {102}},\ \bibinfo {pages} {245410} (\bibinfo {year} {2020})}\BibitemShut
  {NoStop}%
\bibitem [{\citenamefont {Miravet}\ \emph {et~al.}(2023)\citenamefont
  {Miravet}, \citenamefont {Alt\ifmmode \imath \else \i
  \fi{}nta\ifmmode~\mbox{\c{s}}\else \c{s}\fi{}}, \citenamefont {Rodrigues},
  \citenamefont {Bieniek}, \citenamefont {Korkusinski},\ and\ \citenamefont
  {Hawrylak}}]{Daniel2023}%
  \BibitemOpen
  \bibfield  {author} {\bibinfo {author} {\bibfnamefont {D.}~\bibnamefont
  {Miravet}}, \bibinfo {author} {\bibfnamefont {A.}~\bibnamefont {Alt\ifmmode
  \imath \else \i \fi{}nta\ifmmode~\mbox{\c{s}}\else \c{s}\fi{}}}, \bibinfo
  {author} {\bibfnamefont {A.~W.}\ \bibnamefont {Rodrigues}}, \bibinfo {author}
  {\bibfnamefont {M.}~\bibnamefont {Bieniek}}, \bibinfo {author} {\bibfnamefont
  {M.}~\bibnamefont {Korkusinski}},\ and\ \bibinfo {author} {\bibfnamefont
  {P.}~\bibnamefont {Hawrylak}},\ }\bibfield  {title} {\bibinfo {title}
  {Interacting holes in gated {WSe}$_{2}$ quantum dots},\ }\href
  {https://doi.org/10.1103/PhysRevB.108.195407} {\bibfield  {journal} {\bibinfo
   {journal} {Phys. Rev. B}\ }\textbf {\bibinfo {volume} {108}},\ \bibinfo
  {pages} {195407} (\bibinfo {year} {2023})}\BibitemShut {NoStop}%
\bibitem [{\citenamefont {Saleem}\ \emph {et~al.}(2023)\citenamefont {Saleem},
  \citenamefont {Sadecka}, \citenamefont {Korkusinski}, \citenamefont
  {Miravet}, \citenamefont {Dusko},\ and\ \citenamefont
  {Hawrylak}}]{Saleem2023}%
  \BibitemOpen
  \bibfield  {author} {\bibinfo {author} {\bibfnamefont {Y.}~\bibnamefont
  {Saleem}}, \bibinfo {author} {\bibfnamefont {K.}~\bibnamefont {Sadecka}},
  \bibinfo {author} {\bibfnamefont {M.}~\bibnamefont {Korkusinski}}, \bibinfo
  {author} {\bibfnamefont {D.}~\bibnamefont {Miravet}}, \bibinfo {author}
  {\bibfnamefont {A.}~\bibnamefont {Dusko}},\ and\ \bibinfo {author}
  {\bibfnamefont {P.}~\bibnamefont {Hawrylak}},\ }\bibfield  {title} {\bibinfo
  {title} {Theory of excitons in gated bilayer graphene quantum dots},\ }\href
  {https://doi.org/10.1021/acs.nanolett.3c00406} {\bibfield  {journal}
  {\bibinfo  {journal} {Nano Letters}\ }\textbf {\bibinfo {volume} {23}},\
  \bibinfo {pages} {2998} (\bibinfo {year} {2023})}\BibitemShut {NoStop}%
\bibitem [{\citenamefont {Pisoni}\ \emph {et~al.}(2017)\citenamefont {Pisoni},
  \citenamefont {Lee}, \citenamefont {Overweg}, \citenamefont {Eich},
  \citenamefont {Simonet}, \citenamefont {Watanabe}, \citenamefont {Taniguchi},
  \citenamefont {Gorbachev}, \citenamefont {Ihn},\ and\ \citenamefont
  {Ensslin}}]{Pisoni2017}%
  \BibitemOpen
  \bibfield  {author} {\bibinfo {author} {\bibfnamefont {R.}~\bibnamefont
  {Pisoni}}, \bibinfo {author} {\bibfnamefont {Y.}~\bibnamefont {Lee}},
  \bibinfo {author} {\bibfnamefont {H.}~\bibnamefont {Overweg}}, \bibinfo
  {author} {\bibfnamefont {M.}~\bibnamefont {Eich}}, \bibinfo {author}
  {\bibfnamefont {P.}~\bibnamefont {Simonet}}, \bibinfo {author} {\bibfnamefont
  {K.}~\bibnamefont {Watanabe}}, \bibinfo {author} {\bibfnamefont
  {T.}~\bibnamefont {Taniguchi}}, \bibinfo {author} {\bibfnamefont
  {R.}~\bibnamefont {Gorbachev}}, \bibinfo {author} {\bibfnamefont
  {T.}~\bibnamefont {Ihn}},\ and\ \bibinfo {author} {\bibfnamefont
  {K.}~\bibnamefont {Ensslin}},\ }\bibfield  {title} {\bibinfo {title}
  {Gate-defined one-dimensional channel and broken symmetry states in {MoS}$_2$
  van der {Waals} heterostructures},\ }\bibfield  {booktitle} {\emph {\bibinfo
  {booktitle} {Nano Letters}},\ }\href
  {https://doi.org/10.1021/acs.nanolett.7b02186} {\bibfield  {journal}
  {\bibinfo  {journal} {Nano Letters}\ }\textbf {\bibinfo {volume} {17}},\
  \bibinfo {pages} {5008} (\bibinfo {year} {2017})}\BibitemShut {NoStop}%
\bibitem [{\citenamefont {Goossens}\ \emph {et~al.}(2012)\citenamefont
  {Goossens}, \citenamefont {Driessen}, \citenamefont {Baart}, \citenamefont
  {Watanabe}, \citenamefont {Taniguchi},\ and\ \citenamefont
  {Vandersypen}}]{Goossens2012b}%
  \BibitemOpen
  \bibfield  {author} {\bibinfo {author} {\bibfnamefont {A.~S.~M.}\
  \bibnamefont {Goossens}}, \bibinfo {author} {\bibfnamefont {S.~C.~M.}\
  \bibnamefont {Driessen}}, \bibinfo {author} {\bibfnamefont {T.~A.}\
  \bibnamefont {Baart}}, \bibinfo {author} {\bibfnamefont {K.}~\bibnamefont
  {Watanabe}}, \bibinfo {author} {\bibfnamefont {T.}~\bibnamefont
  {Taniguchi}},\ and\ \bibinfo {author} {\bibfnamefont {L.~M.~K.}\ \bibnamefont
  {Vandersypen}},\ }\bibfield  {title} {\bibinfo {title} {Gate-defined
  confinement in bilayer graphene-hexagonal boron nitride hybrid devices},\
  }\href {https://doi.org/10.1021/nl301986q} {\bibfield  {journal} {\bibinfo
  {journal} {Nano Letters}\ }\textbf {\bibinfo {volume} {12}},\ \bibinfo
  {pages} {4656} (\bibinfo {year} {2012})}\BibitemShut {NoStop}%
\bibitem [{\citenamefont {Boddison-Chouinard}\ \emph
  {et~al.}(2023)\citenamefont {Boddison-Chouinard}, \citenamefont {Bogan},
  \citenamefont {Barrios}, \citenamefont {Lapointe}, \citenamefont {Watanabe},
  \citenamefont {Taniguchi}, \citenamefont {Paw{\l}owski}, \citenamefont
  {Miravet}, \citenamefont {Bieniek}, \citenamefont {Hawrylak}, \citenamefont
  {Luican-Mayer},\ and\ \citenamefont {Gaudreau}}]{Justin2023}%
  \BibitemOpen
  \bibfield  {author} {\bibinfo {author} {\bibfnamefont {J.}~\bibnamefont
  {Boddison-Chouinard}}, \bibinfo {author} {\bibfnamefont {A.}~\bibnamefont
  {Bogan}}, \bibinfo {author} {\bibfnamefont {P.}~\bibnamefont {Barrios}},
  \bibinfo {author} {\bibfnamefont {J.}~\bibnamefont {Lapointe}}, \bibinfo
  {author} {\bibfnamefont {K.}~\bibnamefont {Watanabe}}, \bibinfo {author}
  {\bibfnamefont {T.}~\bibnamefont {Taniguchi}}, \bibinfo {author}
  {\bibfnamefont {J.}~\bibnamefont {Paw{\l}owski}}, \bibinfo {author}
  {\bibfnamefont {D.}~\bibnamefont {Miravet}}, \bibinfo {author} {\bibfnamefont
  {M.}~\bibnamefont {Bieniek}}, \bibinfo {author} {\bibfnamefont
  {P.}~\bibnamefont {Hawrylak}}, \bibinfo {author} {\bibfnamefont
  {A.}~\bibnamefont {Luican-Mayer}},\ and\ \bibinfo {author} {\bibfnamefont
  {L.}~\bibnamefont {Gaudreau}},\ }\bibfield  {title} {\bibinfo {title}
  {Anomalous conductance quantization of a one-dimensional channel in monolayer
  {WSe}$_2$},\ }\href {https://doi.org/10.1038/s41699-023-00407-y} {\bibfield
  {journal} {\bibinfo  {journal} {npj 2D Materials and Applications}\ }\textbf
  {\bibinfo {volume} {7}},\ \bibinfo {pages} {50} (\bibinfo {year}
  {2023})}\BibitemShut {NoStop}%
\bibitem [{\citenamefont {Braz}\ \emph {et~al.}(2018)\citenamefont {Braz},
  \citenamefont {Amorim},\ and\ \citenamefont {Castro}}]{Braz_Castro_2018}%
  \BibitemOpen
  \bibfield  {author} {\bibinfo {author} {\bibfnamefont {J.~E.~H.}\
  \bibnamefont {Braz}}, \bibinfo {author} {\bibfnamefont {B.}~\bibnamefont
  {Amorim}},\ and\ \bibinfo {author} {\bibfnamefont {E.~V.}\ \bibnamefont
  {Castro}},\ }\bibfield  {title} {\bibinfo {title} {Valley-polarized magnetic
  state in hole-doped monolayers of transition-metal dichalcogenides},\ }\href
  {https://doi.org/10.1103/PhysRevB.98.161406} {\bibfield  {journal} {\bibinfo
  {journal} {Phys. Rev. B}\ }\textbf {\bibinfo {volume} {98}},\ \bibinfo
  {pages} {161406} (\bibinfo {year} {2018})}\BibitemShut {NoStop}%
\bibitem [{\citenamefont {Van~der Donck}\ and\ \citenamefont
  {Peeters}(2018)}]{Donck_Peeters_2018}%
  \BibitemOpen
  \bibfield  {author} {\bibinfo {author} {\bibfnamefont {M.}~\bibnamefont
  {Van~der Donck}}\ and\ \bibinfo {author} {\bibfnamefont {F.~M.}\ \bibnamefont
  {Peeters}},\ }\bibfield  {title} {\bibinfo {title} {Rich many-body phase
  diagram of electrons and holes in doped monolayer transition metal
  dichalcogenides},\ }\href {https://doi.org/10.1103/PhysRevB.98.115432}
  {\bibfield  {journal} {\bibinfo  {journal} {Phys. Rev. B}\ }\textbf {\bibinfo
  {volume} {98}},\ \bibinfo {pages} {115432} (\bibinfo {year}
  {2018})}\BibitemShut {NoStop}%
\bibitem [{\citenamefont {Attaccalite}\ \emph {et~al.}(2002)\citenamefont
  {Attaccalite}, \citenamefont {Moroni}, \citenamefont {Gori-Giorgi},\ and\
  \citenamefont {Bachelet}}]{Attaccalite2002}%
  \BibitemOpen
  \bibfield  {author} {\bibinfo {author} {\bibfnamefont {C.}~\bibnamefont
  {Attaccalite}}, \bibinfo {author} {\bibfnamefont {S.}~\bibnamefont {Moroni}},
  \bibinfo {author} {\bibfnamefont {P.}~\bibnamefont {Gori-Giorgi}},\ and\
  \bibinfo {author} {\bibfnamefont {G.~B.}\ \bibnamefont {Bachelet}},\
  }\bibfield  {title} {\bibinfo {title} {Correlation energy and spin
  polarization in the {2D} electron gas},\ }\href
  {https://doi.org/10.1103/PhysRevLett.88.256601} {\bibfield  {journal}
  {\bibinfo  {journal} {Phys. Rev. Lett.}\ }\textbf {\bibinfo {volume} {88}},\
  \bibinfo {pages} {256601} (\bibinfo {year} {2002})}\BibitemShut {NoStop}%
\bibitem [{\citenamefont {Suba\ifmmode~\mbox{\c{s}}\else \c{s}\fi{}i}\ and\
  \citenamefont {Tanatar}(2008)}]{Tanatar2008}%
  \BibitemOpen
  \bibfield  {author} {\bibinfo {author} {\bibfnamefont {A.~L.}\ \bibnamefont
  {Suba\ifmmode~\mbox{\c{s}}\else \c{s}\fi{}i}}\ and\ \bibinfo {author}
  {\bibfnamefont {B.}~\bibnamefont {Tanatar}},\ }\bibfield  {title} {\bibinfo
  {title} {Effects of a parallel magnetic field on the ground-state magnetic
  properties of a two-dimensional electron gas},\ }\href
  {https://doi.org/10.1103/PhysRevB.78.155304} {\bibfield  {journal} {\bibinfo
  {journal} {Phys. Rev. B}\ }\textbf {\bibinfo {volume} {78}},\ \bibinfo
  {pages} {155304} (\bibinfo {year} {2008})}\BibitemShut {NoStop}%
\bibitem [{\citenamefont {Meir}\ \emph {et~al.}(2002)\citenamefont {Meir},
  \citenamefont {Hirose},\ and\ \citenamefont {Wingreen}}]{Wingreen2002a}%
  \BibitemOpen
  \bibfield  {author} {\bibinfo {author} {\bibfnamefont {Y.}~\bibnamefont
  {Meir}}, \bibinfo {author} {\bibfnamefont {K.}~\bibnamefont {Hirose}},\ and\
  \bibinfo {author} {\bibfnamefont {N.~S.}\ \bibnamefont {Wingreen}},\
  }\bibfield  {title} {\bibinfo {title} {Kondo model for the ``0.7 anomaly'' in
  transport through a quantum point contact},\ }\href
  {https://doi.org/10.1103/PhysRevLett.89.196802} {\bibfield  {journal}
  {\bibinfo  {journal} {Phys. Rev. Lett.}\ }\textbf {\bibinfo {volume} {89}},\
  \bibinfo {pages} {196802} (\bibinfo {year} {2002})}\BibitemShut {NoStop}%
\bibitem [{\citenamefont {Cronenwett}\ \emph {et~al.}(2002)\citenamefont
  {Cronenwett}, \citenamefont {Lynch}, \citenamefont {Goldhaber-Gordon},
  \citenamefont {Kouwenhoven}, \citenamefont {Marcus}, \citenamefont {Hirose},
  \citenamefont {Wingreen},\ and\ \citenamefont {Umansky}}]{Wingreen2002b}%
  \BibitemOpen
  \bibfield  {author} {\bibinfo {author} {\bibfnamefont {S.~M.}\ \bibnamefont
  {Cronenwett}}, \bibinfo {author} {\bibfnamefont {H.~J.}\ \bibnamefont
  {Lynch}}, \bibinfo {author} {\bibfnamefont {D.}~\bibnamefont
  {Goldhaber-Gordon}}, \bibinfo {author} {\bibfnamefont {L.~P.}\ \bibnamefont
  {Kouwenhoven}}, \bibinfo {author} {\bibfnamefont {C.~M.}\ \bibnamefont
  {Marcus}}, \bibinfo {author} {\bibfnamefont {K.}~\bibnamefont {Hirose}},
  \bibinfo {author} {\bibfnamefont {N.~S.}\ \bibnamefont {Wingreen}},\ and\
  \bibinfo {author} {\bibfnamefont {V.}~\bibnamefont {Umansky}},\ }\bibfield
  {title} {\bibinfo {title} {Low-temperature fate of the 0.7 structure in a
  point contact: A {Kondo}-like correlated state in an open system},\ }\href
  {https://doi.org/10.1103/PhysRevLett.88.226805} {\bibfield  {journal}
  {\bibinfo  {journal} {Phys. Rev. Lett.}\ }\textbf {\bibinfo {volume} {88}},\
  \bibinfo {pages} {226805} (\bibinfo {year} {2002})}\BibitemShut {NoStop}%
\bibitem [{\citenamefont {G\"u\ifmmode~\mbox{\c{c}}\else \c{c}\fi{}l\"u}\ \emph
  {et~al.}(2009{\natexlab{b}})\citenamefont {G\"u\ifmmode~\mbox{\c{c}}\else
  \c{c}\fi{}l\"u}, \citenamefont {Umrigar}, \citenamefont {Jiang},\ and\
  \citenamefont {Baranger}}]{Guclu_Baranger_2009}%
  \BibitemOpen
  \bibfield  {author} {\bibinfo {author} {\bibfnamefont {A.~D.}\ \bibnamefont
  {G\"u\ifmmode~\mbox{\c{c}}\else \c{c}\fi{}l\"u}}, \bibinfo {author}
  {\bibfnamefont {C.~J.}\ \bibnamefont {Umrigar}}, \bibinfo {author}
  {\bibfnamefont {H.}~\bibnamefont {Jiang}},\ and\ \bibinfo {author}
  {\bibfnamefont {H.~U.}\ \bibnamefont {Baranger}},\ }\bibfield  {title}
  {\bibinfo {title} {Localization in an inhomogeneous quantum wire},\ }\href
  {https://doi.org/10.1103/PhysRevB.80.201302} {\bibfield  {journal} {\bibinfo
  {journal} {Phys. Rev. B}\ }\textbf {\bibinfo {volume} {80}},\ \bibinfo
  {pages} {201302} (\bibinfo {year} {2009}{\natexlab{b}})}\BibitemShut
  {NoStop}%
\bibitem [{\citenamefont {Giuliani}\ and\ \citenamefont
  {Vignale}(2008)}]{giuliani2008quantum}%
  \BibitemOpen
  \bibfield  {author} {\bibinfo {author} {\bibfnamefont {G.}~\bibnamefont
  {Giuliani}}\ and\ \bibinfo {author} {\bibfnamefont {G.}~\bibnamefont
  {Vignale}},\ }\href@noop {} {\emph {\bibinfo {title} {Quantum theory of the
  electron liquid}}}\ (\bibinfo  {publisher} {Cambridge university press},\
  \bibinfo {year} {2008})\BibitemShut {NoStop}%
\bibitem [{\citenamefont {Matveev}(2004)}]{Matveev_2004}%
  \BibitemOpen
  \bibfield  {author} {\bibinfo {author} {\bibfnamefont {K.~A.}\ \bibnamefont
  {Matveev}},\ }\bibfield  {title} {\bibinfo {title} {Conductance of a quantum
  wire in the {Wigner}-crystal regime},\ }\href
  {https://doi.org/10.1103/PhysRevLett.92.106801} {\bibfield  {journal}
  {\bibinfo  {journal} {Phys. Rev. Lett.}\ }\textbf {\bibinfo {volume} {92}},\
  \bibinfo {pages} {106801} (\bibinfo {year} {2004})}\BibitemShut {NoStop}%
\bibitem [{\citenamefont {Bieniek}\ \emph {et~al.}(2018)\citenamefont
  {Bieniek}, \citenamefont {Korkusi\ifmmode~\acute{n}\else \'{n}\fi{}ski},
  \citenamefont {Szulakowska}, \citenamefont {Potasz}, \citenamefont
  {Ozfidan},\ and\ \citenamefont {Hawrylak}}]{Bieniek_Hawrylak_2018}%
  \BibitemOpen
  \bibfield  {author} {\bibinfo {author} {\bibfnamefont {M.}~\bibnamefont
  {Bieniek}}, \bibinfo {author} {\bibfnamefont {M.}~\bibnamefont
  {Korkusi\ifmmode~\acute{n}\else \'{n}\fi{}ski}}, \bibinfo {author}
  {\bibfnamefont {L.}~\bibnamefont {Szulakowska}}, \bibinfo {author}
  {\bibfnamefont {P.}~\bibnamefont {Potasz}}, \bibinfo {author} {\bibfnamefont
  {I.}~\bibnamefont {Ozfidan}},\ and\ \bibinfo {author} {\bibfnamefont
  {P.}~\bibnamefont {Hawrylak}},\ }\bibfield  {title} {\bibinfo {title} {Band
  nesting, massive {Dirac} fermions, and valley {Land\'e} and {Zeeman} effects
  in transition metal dichalcogenides: A tight-binding model},\ }\href
  {https://doi.org/10.1103/PhysRevB.97.085153} {\bibfield  {journal} {\bibinfo
  {journal} {Phys. Rev. B}\ }\textbf {\bibinfo {volume} {97}},\ \bibinfo
  {pages} {085153} (\bibinfo {year} {2018})}\BibitemShut {NoStop}%
\bibitem [{\citenamefont {Bieniek}\ \emph {et~al.}(2020)\citenamefont
  {Bieniek}, \citenamefont {Szulakowska},\ and\ \citenamefont
  {Hawrylak}}]{Bieniek_Hawrylak_2020}%
  \BibitemOpen
  \bibfield  {author} {\bibinfo {author} {\bibfnamefont {M.}~\bibnamefont
  {Bieniek}}, \bibinfo {author} {\bibfnamefont {L.}~\bibnamefont
  {Szulakowska}},\ and\ \bibinfo {author} {\bibfnamefont {P.}~\bibnamefont
  {Hawrylak}},\ }\bibfield  {title} {\bibinfo {title} {Effect of valley, spin,
  and band nesting on the electronic properties of gated quantum dots in a
  single layer of transition metal dichalcogenides},\ }\href
  {https://doi.org/10.1103/PhysRevB.101.035401} {\bibfield  {journal} {\bibinfo
   {journal} {Phys. Rev. B}\ }\textbf {\bibinfo {volume} {101}},\ \bibinfo
  {pages} {035401} (\bibinfo {year} {2020})}\BibitemShut {NoStop}%
\bibitem [{\citenamefont {Roldán}\ \emph {et~al.}(2014)\citenamefont
  {Roldán}, \citenamefont {López-Sancho}, \citenamefont {Guinea},
  \citenamefont {Cappelluti}, \citenamefont {Silva-Guillén},\ and\
  \citenamefont {Ordejón}}]{Roldan2014}%
  \BibitemOpen
  \bibfield  {author} {\bibinfo {author} {\bibfnamefont {R.}~\bibnamefont
  {Roldán}}, \bibinfo {author} {\bibfnamefont {M.~P.}\ \bibnamefont
  {López-Sancho}}, \bibinfo {author} {\bibfnamefont {F.}~\bibnamefont
  {Guinea}}, \bibinfo {author} {\bibfnamefont {E.}~\bibnamefont {Cappelluti}},
  \bibinfo {author} {\bibfnamefont {J.~A.}\ \bibnamefont {Silva-Guillén}},\
  and\ \bibinfo {author} {\bibfnamefont {P.}~\bibnamefont {Ordejón}},\
  }\bibfield  {title} {\bibinfo {title} {Momentum dependence of spin–orbit
  interaction effects in single-layer and multi-layer transition metal
  dichalcogenides},\ }\href {https://doi.org/10.1088/2053-1583/1/3/034003}
  {\bibfield  {journal} {\bibinfo  {journal} {2D Materials}\ }\textbf {\bibinfo
  {volume} {1}},\ \bibinfo {pages} {034003} (\bibinfo {year}
  {2014})}\BibitemShut {NoStop}%
\bibitem [{\citenamefont {Goldberg}\ \emph {et~al.}(2024)\citenamefont
  {Goldberg}, \citenamefont {Yannouleas},\ and\ \citenamefont
  {Landman}}]{Goldberg2024}%
  \BibitemOpen
  \bibfield  {author} {\bibinfo {author} {\bibfnamefont {A.}~\bibnamefont
  {Goldberg}}, \bibinfo {author} {\bibfnamefont {C.}~\bibnamefont
  {Yannouleas}},\ and\ \bibinfo {author} {\bibfnamefont {U.}~\bibnamefont
  {Landman}},\ }\bibfield  {title} {\bibinfo {title} {Electronic
  wigner-molecule polymeric chains in elongated silicon quantum dots and
  finite-length quantum wires},\ }\href
  {https://doi.org/10.1103/PhysRevApplied.21.064063} {\bibfield  {journal}
  {\bibinfo  {journal} {Phys. Rev. Appl.}\ }\textbf {\bibinfo {volume} {21}},\
  \bibinfo {pages} {064063} (\bibinfo {year} {2024})}\BibitemShut {NoStop}%
\bibitem [{\citenamefont {Winkler}(2003)}]{Winkler}%
  \BibitemOpen
  \bibfield  {author} {\bibinfo {author} {\bibfnamefont {R.}~\bibnamefont
  {Winkler}},\ }\href {https://doi.org/10.1007/b13586} {\emph {\bibinfo {title}
  {Spin—Orbit Coupling Effects in Two-Dimensional Electron and Hole
  Systems}}}\ (\bibinfo  {publisher} {Springer Berlin Heidelberg},\ \bibinfo
  {year} {2003})\BibitemShut {NoStop}%
\bibitem [{\citenamefont {Paw\l{}owski}\ \emph {et~al.}(2016)\citenamefont
  {Paw\l{}owski}, \citenamefont {Szumniak},\ and\ \citenamefont
  {Bednarek}}]{pawlowski_2016}%
  \BibitemOpen
  \bibfield  {author} {\bibinfo {author} {\bibfnamefont {J.}~\bibnamefont
  {Paw\l{}owski}}, \bibinfo {author} {\bibfnamefont {P.}~\bibnamefont
  {Szumniak}},\ and\ \bibinfo {author} {\bibfnamefont {S.}~\bibnamefont
  {Bednarek}},\ }\bibfield  {title} {\bibinfo {title} {Generation of
  spin-dependent coherent states in a quantum wire},\ }\href
  {https://doi.org/10.1103/PhysRevB.94.155407} {\bibfield  {journal} {\bibinfo
  {journal} {Phys. Rev. B}\ }\textbf {\bibinfo {volume} {94}},\ \bibinfo
  {pages} {155407} (\bibinfo {year} {2016})}\BibitemShut {NoStop}%
\bibitem [{\citenamefont {Korm\'anyos}\ \emph {et~al.}(2014)\citenamefont
  {Korm\'anyos}, \citenamefont {Z\'olyomi}, \citenamefont {Drummond},\ and\
  \citenamefont {Burkard}}]{Kormanyos}%
  \BibitemOpen
  \bibfield  {author} {\bibinfo {author} {\bibfnamefont {A.}~\bibnamefont
  {Korm\'anyos}}, \bibinfo {author} {\bibfnamefont {V.}~\bibnamefont
  {Z\'olyomi}}, \bibinfo {author} {\bibfnamefont {N.~D.}\ \bibnamefont
  {Drummond}},\ and\ \bibinfo {author} {\bibfnamefont {G.}~\bibnamefont
  {Burkard}},\ }\bibfield  {title} {\bibinfo {title} {Spin-orbit coupling,
  quantum dots, and qubits in monolayer transition metal dichalcogenides},\
  }\href {https://doi.org/10.1103/PhysRevX.4.011034} {\bibfield  {journal}
  {\bibinfo  {journal} {Phys. Rev. X}\ }\textbf {\bibinfo {volume} {4}},\
  \bibinfo {pages} {011034} (\bibinfo {year} {2014})}\BibitemShut {NoStop}%
\bibitem [{\citenamefont {Paw\l{}owski}\ \emph {et~al.}(2017)\citenamefont
  {Paw\l{}owski}, \citenamefont {G\'orski}, \citenamefont {Skowron},\ and\
  \citenamefont {Bednarek}}]{Catstates}%
  \BibitemOpen
  \bibfield  {author} {\bibinfo {author} {\bibfnamefont {J.}~\bibnamefont
  {Paw\l{}owski}}, \bibinfo {author} {\bibfnamefont {M.}~\bibnamefont
  {G\'orski}}, \bibinfo {author} {\bibfnamefont {G.}~\bibnamefont {Skowron}},\
  and\ \bibinfo {author} {\bibfnamefont {S.}~\bibnamefont {Bednarek}},\
  }\bibfield  {title} {\bibinfo {title} {Generation of {Schr\"odinger} cat type
  states in a planar semiconductor heterostructure},\ }\href
  {https://doi.org/10.1103/PhysRevB.96.115308} {\bibfield  {journal} {\bibinfo
  {journal} {Phys. Rev. B}\ }\textbf {\bibinfo {volume} {96}},\ \bibinfo
  {pages} {115308} (\bibinfo {year} {2017})}\BibitemShut {NoStop}%
\bibitem [{\citenamefont {Rapisarda}\ and\ \citenamefont
  {Senatore}(1996)}]{Rapisarda1996}%
  \BibitemOpen
  \bibfield  {author} {\bibinfo {author} {\bibfnamefont {F.}~\bibnamefont
  {Rapisarda}}\ and\ \bibinfo {author} {\bibfnamefont {G.}~\bibnamefont
  {Senatore}},\ }\bibfield  {title} {\bibinfo {title} {Diffusion {Monte Carlo}
  study of electrons in two-dimensional layers},\ }\href
  {https://doi.org/10.1071/PH960161} {\bibfield  {journal} {\bibinfo  {journal}
  {Australian Journal of Physics}\ }\textbf {\bibinfo {volume} {49}},\ \bibinfo
  {pages} {161} (\bibinfo {year} {1996})}\BibitemShut {NoStop}%
\bibitem [{\citenamefont {Ortiz}\ and\ \citenamefont
  {Ballone}(1994)}]{Ortiz1994}%
  \BibitemOpen
  \bibfield  {author} {\bibinfo {author} {\bibfnamefont {G.}~\bibnamefont
  {Ortiz}}\ and\ \bibinfo {author} {\bibfnamefont {P.}~\bibnamefont
  {Ballone}},\ }\bibfield  {title} {\bibinfo {title} {Correlation energy,
  structure factor, radial distribution function, and momentum distribution of
  the spin-polarized uniform electron gas},\ }\href
  {https://doi.org/10.1103/PhysRevB.50.1391} {\bibfield  {journal} {\bibinfo
  {journal} {Phys. Rev. B}\ }\textbf {\bibinfo {volume} {50}},\ \bibinfo
  {pages} {1391} (\bibinfo {year} {1994})}\BibitemShut {NoStop}%
\bibitem [{\citenamefont {G\"u\ifmmode~\mbox{\c{c}}\else
  \c{c}\fi{}l\"u}(2016)}]{Guclu2016}%
  \BibitemOpen
  \bibfield  {author} {\bibinfo {author} {\bibfnamefont {A.~D.}\ \bibnamefont
  {G\"u\ifmmode~\mbox{\c{c}}\else \c{c}\fi{}l\"u}},\ }\bibfield  {title}
  {\bibinfo {title} {Wigner crystallization at graphene edges},\ }\href
  {https://doi.org/10.1103/PhysRevB.93.045114} {\bibfield  {journal} {\bibinfo
  {journal} {Phys. Rev. B}\ }\textbf {\bibinfo {volume} {93}},\ \bibinfo
  {pages} {045114} (\bibinfo {year} {2016})}\BibitemShut {NoStop}%
\bibitem [{\citenamefont {Meyer}\ and\ \citenamefont
  {Matveev}(2008)}]{Meyer_2009}%
  \BibitemOpen
  \bibfield  {author} {\bibinfo {author} {\bibfnamefont {J.~S.}\ \bibnamefont
  {Meyer}}\ and\ \bibinfo {author} {\bibfnamefont {K.~A.}\ \bibnamefont
  {Matveev}},\ }\bibfield  {title} {\bibinfo {title} {Wigner crystal physics in
  quantum wires},\ }\href {https://doi.org/10.1088/0953-8984/21/2/023203}
  {\bibfield  {journal} {\bibinfo  {journal} {Journal of Physics: Condensed
  Matter}\ }\textbf {\bibinfo {volume} {21}},\ \bibinfo {pages} {023203}
  (\bibinfo {year} {2008})}\BibitemShut {NoStop}%
\bibitem [{\citenamefont {Piacente}\ \emph {et~al.}(2004)\citenamefont
  {Piacente}, \citenamefont {Schweigert}, \citenamefont {Betouras},\ and\
  \citenamefont {Peeters}}]{Piacente2004}%
  \BibitemOpen
  \bibfield  {author} {\bibinfo {author} {\bibfnamefont {G.}~\bibnamefont
  {Piacente}}, \bibinfo {author} {\bibfnamefont {I.~V.}\ \bibnamefont
  {Schweigert}}, \bibinfo {author} {\bibfnamefont {J.~J.}\ \bibnamefont
  {Betouras}},\ and\ \bibinfo {author} {\bibfnamefont {F.~M.}\ \bibnamefont
  {Peeters}},\ }\bibfield  {title} {\bibinfo {title} {Generic properties of a
  quasi-one-dimensional classical {Wigner} crystal},\ }\href
  {https://doi.org/10.1103/PhysRevB.69.045324} {\bibfield  {journal} {\bibinfo
  {journal} {Phys. Rev. B}\ }\textbf {\bibinfo {volume} {69}},\ \bibinfo
  {pages} {045324} (\bibinfo {year} {2004})}\BibitemShut {NoStop}%
\bibitem [{\citenamefont {Klironomos}\ \emph {et~al.}(2007)\citenamefont
  {Klironomos}, \citenamefont {Meyer}, \citenamefont {Hikihara},\ and\
  \citenamefont {Matveev}}]{Klironomos2007}%
  \BibitemOpen
  \bibfield  {author} {\bibinfo {author} {\bibfnamefont {A.~D.}\ \bibnamefont
  {Klironomos}}, \bibinfo {author} {\bibfnamefont {J.~S.}\ \bibnamefont
  {Meyer}}, \bibinfo {author} {\bibfnamefont {T.}~\bibnamefont {Hikihara}},\
  and\ \bibinfo {author} {\bibfnamefont {K.~A.}\ \bibnamefont {Matveev}},\
  }\bibfield  {title} {\bibinfo {title} {Spin coupling in zigzag {Wigner}
  crystals},\ }\href {https://doi.org/10.1103/PhysRevB.76.075302} {\bibfield
  {journal} {\bibinfo  {journal} {Phys. Rev. B}\ }\textbf {\bibinfo {volume}
  {76}},\ \bibinfo {pages} {075302} (\bibinfo {year} {2007})}\BibitemShut
  {NoStop}%
\bibitem [{\citenamefont {Mehta}\ \emph {et~al.}(2013)\citenamefont {Mehta},
  \citenamefont {Umrigar}, \citenamefont {Meyer},\ and\ \citenamefont
  {Baranger}}]{Mehta_Baranger_2013}%
  \BibitemOpen
  \bibfield  {author} {\bibinfo {author} {\bibfnamefont {A.~C.}\ \bibnamefont
  {Mehta}}, \bibinfo {author} {\bibfnamefont {C.~J.}\ \bibnamefont {Umrigar}},
  \bibinfo {author} {\bibfnamefont {J.~S.}\ \bibnamefont {Meyer}},\ and\
  \bibinfo {author} {\bibfnamefont {H.~U.}\ \bibnamefont {Baranger}},\
  }\bibfield  {title} {\bibinfo {title} {Zigzag phase transition in quantum
  wires},\ }\href {https://doi.org/10.1103/PhysRevLett.110.246802} {\bibfield
  {journal} {\bibinfo  {journal} {Phys. Rev. Lett.}\ }\textbf {\bibinfo
  {volume} {110}},\ \bibinfo {pages} {246802} (\bibinfo {year}
  {2013})}\BibitemShut {NoStop}%
\bibitem [{\citenamefont {Zhou}\ \emph {et~al.}(2021)\citenamefont {Zhou},
  \citenamefont {Sung}, \citenamefont {Brutschea}, \citenamefont {Esterlis},
  \citenamefont {Wang}, \citenamefont {Scuri}, \citenamefont {Gelly},
  \citenamefont {Heo}, \citenamefont {Taniguchi}, \citenamefont {Watanabe},
  \citenamefont {Zar{\'a}nd}, \citenamefont {Lukin}, \citenamefont {Kim},
  \citenamefont {Demler},\ and\ \citenamefont {Park}}]{Zhou2021}%
  \BibitemOpen
  \bibfield  {author} {\bibinfo {author} {\bibfnamefont {Y.}~\bibnamefont
  {Zhou}}, \bibinfo {author} {\bibfnamefont {J.}~\bibnamefont {Sung}}, \bibinfo
  {author} {\bibfnamefont {E.}~\bibnamefont {Brutschea}}, \bibinfo {author}
  {\bibfnamefont {I.}~\bibnamefont {Esterlis}}, \bibinfo {author}
  {\bibfnamefont {Y.}~\bibnamefont {Wang}}, \bibinfo {author} {\bibfnamefont
  {G.}~\bibnamefont {Scuri}}, \bibinfo {author} {\bibfnamefont {R.~J.}\
  \bibnamefont {Gelly}}, \bibinfo {author} {\bibfnamefont {H.}~\bibnamefont
  {Heo}}, \bibinfo {author} {\bibfnamefont {T.}~\bibnamefont {Taniguchi}},
  \bibinfo {author} {\bibfnamefont {K.}~\bibnamefont {Watanabe}}, \bibinfo
  {author} {\bibfnamefont {G.}~\bibnamefont {Zar{\'a}nd}}, \bibinfo {author}
  {\bibfnamefont {M.~D.}\ \bibnamefont {Lukin}}, \bibinfo {author}
  {\bibfnamefont {P.}~\bibnamefont {Kim}}, \bibinfo {author} {\bibfnamefont
  {E.}~\bibnamefont {Demler}},\ and\ \bibinfo {author} {\bibfnamefont
  {H.}~\bibnamefont {Park}},\ }\bibfield  {title} {\bibinfo {title} {Bilayer
  {Wigner} crystals in a transition metal dichalcogenide heterostructure},\
  }\href {https://doi.org/10.1038/s41586-021-03560-w} {\bibfield  {journal}
  {\bibinfo  {journal} {Nature}\ }\textbf {\bibinfo {volume} {595}},\ \bibinfo
  {pages} {48} (\bibinfo {year} {2021})}\BibitemShut {NoStop}%
\bibitem [{\citenamefont {Li}\ \emph {et~al.}(2021)\citenamefont {Li},
  \citenamefont {Li}, \citenamefont {Regan}, \citenamefont {Wang},
  \citenamefont {Zhao}, \citenamefont {Kahn}, \citenamefont {Yumigeta},
  \citenamefont {Blei}, \citenamefont {Taniguchi}, \citenamefont {Watanabe},
  \citenamefont {Tongay}, \citenamefont {Zettl}, \citenamefont {Crommie},\ and\
  \citenamefont {Wang}}]{Hongyuan2021}%
  \BibitemOpen
  \bibfield  {author} {\bibinfo {author} {\bibfnamefont {H.}~\bibnamefont
  {Li}}, \bibinfo {author} {\bibfnamefont {S.}~\bibnamefont {Li}}, \bibinfo
  {author} {\bibfnamefont {E.~C.}\ \bibnamefont {Regan}}, \bibinfo {author}
  {\bibfnamefont {D.}~\bibnamefont {Wang}}, \bibinfo {author} {\bibfnamefont
  {W.}~\bibnamefont {Zhao}}, \bibinfo {author} {\bibfnamefont {S.}~\bibnamefont
  {Kahn}}, \bibinfo {author} {\bibfnamefont {K.}~\bibnamefont {Yumigeta}},
  \bibinfo {author} {\bibfnamefont {M.}~\bibnamefont {Blei}}, \bibinfo {author}
  {\bibfnamefont {T.}~\bibnamefont {Taniguchi}}, \bibinfo {author}
  {\bibfnamefont {K.}~\bibnamefont {Watanabe}}, \bibinfo {author}
  {\bibfnamefont {S.}~\bibnamefont {Tongay}}, \bibinfo {author} {\bibfnamefont
  {A.}~\bibnamefont {Zettl}}, \bibinfo {author} {\bibfnamefont {M.~F.}\
  \bibnamefont {Crommie}},\ and\ \bibinfo {author} {\bibfnamefont
  {F.}~\bibnamefont {Wang}},\ }\bibfield  {title} {\bibinfo {title} {Imaging
  two-dimensional generalized {Wigner} crystals},\ }\href
  {https://doi.org/10.1038/s41586-021-03874-9} {\bibfield  {journal} {\bibinfo
  {journal} {Nature}\ }\textbf {\bibinfo {volume} {597}},\ \bibinfo {pages}
  {650} (\bibinfo {year} {2021})}\BibitemShut {NoStop}%
\bibitem [{\citenamefont {Li}\ \emph {et~al.}(2024{\natexlab{a}})\citenamefont
  {Li}, \citenamefont {Xiang}, \citenamefont {Regan}, \citenamefont {Zhao},
  \citenamefont {Sailus}, \citenamefont {Banerjee}, \citenamefont {Taniguchi},
  \citenamefont {Watanabe}, \citenamefont {Tongay}, \citenamefont {Zettl},
  \citenamefont {Crommie},\ and\ \citenamefont {Wang}}]{Li2024a}%
  \BibitemOpen
  \bibfield  {author} {\bibinfo {author} {\bibfnamefont {H.}~\bibnamefont
  {Li}}, \bibinfo {author} {\bibfnamefont {Z.}~\bibnamefont {Xiang}}, \bibinfo
  {author} {\bibfnamefont {E.}~\bibnamefont {Regan}}, \bibinfo {author}
  {\bibfnamefont {W.}~\bibnamefont {Zhao}}, \bibinfo {author} {\bibfnamefont
  {R.}~\bibnamefont {Sailus}}, \bibinfo {author} {\bibfnamefont
  {R.}~\bibnamefont {Banerjee}}, \bibinfo {author} {\bibfnamefont
  {T.}~\bibnamefont {Taniguchi}}, \bibinfo {author} {\bibfnamefont
  {K.}~\bibnamefont {Watanabe}}, \bibinfo {author} {\bibfnamefont
  {S.}~\bibnamefont {Tongay}}, \bibinfo {author} {\bibfnamefont
  {A.}~\bibnamefont {Zettl}}, \bibinfo {author} {\bibfnamefont {M.~F.}\
  \bibnamefont {Crommie}},\ and\ \bibinfo {author} {\bibfnamefont
  {F.}~\bibnamefont {Wang}},\ }\bibfield  {title} {\bibinfo {title} {Mapping
  charge excitations in generalized {Wigner} crystals},\ }\href
  {https://doi.org/10.1038/s41565-023-01594-x} {\bibfield  {journal} {\bibinfo
  {journal} {Nature Nanotechnology}\ }\textbf {\bibinfo {volume} {19}},\
  \bibinfo {pages} {618} (\bibinfo {year} {2024}{\natexlab{a}})}\BibitemShut
  {NoStop}%
\bibitem [{\citenamefont {Li}\ \emph {et~al.}(2024{\natexlab{b}})\citenamefont
  {Li}, \citenamefont {Xiang}, \citenamefont {Reddy}, \citenamefont {Devakul},
  \citenamefont {Sailus}, \citenamefont {Banerjee}, \citenamefont {Taniguchi},
  \citenamefont {Watanabe}, \citenamefont {Tongay}, \citenamefont {Zettl},
  \citenamefont {Fu}, \citenamefont {Crommie},\ and\ \citenamefont
  {Wang}}]{li2024b}%
  \BibitemOpen
  \bibfield  {author} {\bibinfo {author} {\bibfnamefont {H.}~\bibnamefont
  {Li}}, \bibinfo {author} {\bibfnamefont {Z.}~\bibnamefont {Xiang}}, \bibinfo
  {author} {\bibfnamefont {A.~P.}\ \bibnamefont {Reddy}}, \bibinfo {author}
  {\bibfnamefont {T.}~\bibnamefont {Devakul}}, \bibinfo {author} {\bibfnamefont
  {R.}~\bibnamefont {Sailus}}, \bibinfo {author} {\bibfnamefont
  {R.}~\bibnamefont {Banerjee}}, \bibinfo {author} {\bibfnamefont
  {T.}~\bibnamefont {Taniguchi}}, \bibinfo {author} {\bibfnamefont
  {K.}~\bibnamefont {Watanabe}}, \bibinfo {author} {\bibfnamefont
  {S.}~\bibnamefont {Tongay}}, \bibinfo {author} {\bibfnamefont
  {A.}~\bibnamefont {Zettl}}, \bibinfo {author} {\bibfnamefont
  {L.}~\bibnamefont {Fu}}, \bibinfo {author} {\bibfnamefont {M.~F.}\
  \bibnamefont {Crommie}},\ and\ \bibinfo {author} {\bibfnamefont
  {F.}~\bibnamefont {Wang}},\ }\bibfield  {title} {\bibinfo {title} {Wigner
  molecular crystals from multielectron moiré artificial atoms},\ }\href
  {https://doi.org/10.1126/science.adk1348} {\bibfield  {journal} {\bibinfo
  {journal} {Science}\ }\textbf {\bibinfo {volume} {385}},\ \bibinfo {pages}
  {86} (\bibinfo {year} {2024}{\natexlab{b}})}\BibitemShut {NoStop}%
\bibitem [{\citenamefont {Li}\ \emph {et~al.}(2024{\natexlab{c}})\citenamefont
  {Li}, \citenamefont {Xiang}, \citenamefont {Wang}, \citenamefont {Naik},
  \citenamefont {Kim}, \citenamefont {Nie}, \citenamefont {Li}, \citenamefont
  {Ge}, \citenamefont {He}, \citenamefont {Ou}, \citenamefont {Banerjee},
  \citenamefont {Taniguchi}, \citenamefont {Watanabe}, \citenamefont {Tongay},
  \citenamefont {Zettl}, \citenamefont {Louie}, \citenamefont {Zaletel},
  \citenamefont {Crommie},\ and\ \citenamefont {Wang}}]{li2024c}%
  \BibitemOpen
  \bibfield  {author} {\bibinfo {author} {\bibfnamefont {H.}~\bibnamefont
  {Li}}, \bibinfo {author} {\bibfnamefont {Z.}~\bibnamefont {Xiang}}, \bibinfo
  {author} {\bibfnamefont {T.}~\bibnamefont {Wang}}, \bibinfo {author}
  {\bibfnamefont {M.~H.}\ \bibnamefont {Naik}}, \bibinfo {author}
  {\bibfnamefont {W.}~\bibnamefont {Kim}}, \bibinfo {author} {\bibfnamefont
  {J.}~\bibnamefont {Nie}}, \bibinfo {author} {\bibfnamefont {S.}~\bibnamefont
  {Li}}, \bibinfo {author} {\bibfnamefont {Z.}~\bibnamefont {Ge}}, \bibinfo
  {author} {\bibfnamefont {Z.}~\bibnamefont {He}}, \bibinfo {author}
  {\bibfnamefont {Y.}~\bibnamefont {Ou}}, \bibinfo {author} {\bibfnamefont
  {R.}~\bibnamefont {Banerjee}}, \bibinfo {author} {\bibfnamefont
  {T.}~\bibnamefont {Taniguchi}}, \bibinfo {author} {\bibfnamefont
  {K.}~\bibnamefont {Watanabe}}, \bibinfo {author} {\bibfnamefont
  {S.}~\bibnamefont {Tongay}}, \bibinfo {author} {\bibfnamefont
  {A.}~\bibnamefont {Zettl}}, \bibinfo {author} {\bibfnamefont {S.~G.}\
  \bibnamefont {Louie}}, \bibinfo {author} {\bibfnamefont {M.~P.}\ \bibnamefont
  {Zaletel}}, \bibinfo {author} {\bibfnamefont {M.~F.}\ \bibnamefont
  {Crommie}},\ and\ \bibinfo {author} {\bibfnamefont {F.}~\bibnamefont
  {Wang}},\ }\bibfield  {title} {\bibinfo {title} {Imaging tunable luttinger
  liquid systems in van der {Waals} heterostructures},\ }\href
  {https://doi.org/10.1038/s41586-024-07596-6} {\bibfield  {journal} {\bibinfo
  {journal} {Nature}\ }\textbf {\bibinfo {volume} {631}},\ \bibinfo {pages}
  {765} (\bibinfo {year} {2024}{\natexlab{c}})}\BibitemShut {NoStop}%
\bibitem [{\citenamefont {Wang}\ \emph {et~al.}(2012)\citenamefont {Wang},
  \citenamefont {Li}, \citenamefont {Chen}, \citenamefont {Xianlong},
  \citenamefont {Rontani},\ and\ \citenamefont {Polini}}]{Wang2012}%
  \BibitemOpen
  \bibfield  {author} {\bibinfo {author} {\bibfnamefont {J.-J.}\ \bibnamefont
  {Wang}}, \bibinfo {author} {\bibfnamefont {W.}~\bibnamefont {Li}}, \bibinfo
  {author} {\bibfnamefont {S.}~\bibnamefont {Chen}}, \bibinfo {author}
  {\bibfnamefont {G.}~\bibnamefont {Xianlong}}, \bibinfo {author}
  {\bibfnamefont {M.}~\bibnamefont {Rontani}},\ and\ \bibinfo {author}
  {\bibfnamefont {M.}~\bibnamefont {Polini}},\ }\bibfield  {title} {\bibinfo
  {title} {Absence of {Wigner} molecules in one-dimensional few-fermion systems
  with short-range interactions},\ }\href
  {https://doi.org/10.1103/PhysRevB.86.075110} {\bibfield  {journal} {\bibinfo
  {journal} {Phys. Rev. B}\ }\textbf {\bibinfo {volume} {86}},\ \bibinfo
  {pages} {075110} (\bibinfo {year} {2012})}\BibitemShut {NoStop}%
\bibitem [{\citenamefont {Kumar}\ \emph
  {et~al.}(2019{\natexlab{a}})\citenamefont {Kumar}, \citenamefont {Pepper},
  \citenamefont {Holmes}, \citenamefont {Montagu}, \citenamefont {Gul},
  \citenamefont {Ritchie},\ and\ \citenamefont {Farrer}}]{Kumar2019}%
  \BibitemOpen
  \bibfield  {author} {\bibinfo {author} {\bibfnamefont {S.}~\bibnamefont
  {Kumar}}, \bibinfo {author} {\bibfnamefont {M.}~\bibnamefont {Pepper}},
  \bibinfo {author} {\bibfnamefont {S.~N.}\ \bibnamefont {Holmes}}, \bibinfo
  {author} {\bibfnamefont {H.}~\bibnamefont {Montagu}}, \bibinfo {author}
  {\bibfnamefont {Y.}~\bibnamefont {Gul}}, \bibinfo {author} {\bibfnamefont
  {D.~A.}\ \bibnamefont {Ritchie}},\ and\ \bibinfo {author} {\bibfnamefont
  {I.}~\bibnamefont {Farrer}},\ }\bibfield  {title} {\bibinfo {title}
  {Zero-magnetic field fractional quantum states},\ }\href
  {https://doi.org/10.1103/PhysRevLett.122.086803} {\bibfield  {journal}
  {\bibinfo  {journal} {Phys. Rev. Lett.}\ }\textbf {\bibinfo {volume} {122}},\
  \bibinfo {pages} {086803} (\bibinfo {year} {2019}{\natexlab{a}})}\BibitemShut
  {NoStop}%
\bibitem [{\citenamefont {Kumar}\ \emph
  {et~al.}(2019{\natexlab{b}})\citenamefont {Kumar}, \citenamefont {Pepper},
  \citenamefont {Ritchie}, \citenamefont {Farrer},\ and\ \citenamefont
  {Montagu}}]{Kumar2019A}%
  \BibitemOpen
  \bibfield  {author} {\bibinfo {author} {\bibfnamefont {S.}~\bibnamefont
  {Kumar}}, \bibinfo {author} {\bibfnamefont {M.}~\bibnamefont {Pepper}},
  \bibinfo {author} {\bibfnamefont {D.}~\bibnamefont {Ritchie}}, \bibinfo
  {author} {\bibfnamefont {I.}~\bibnamefont {Farrer}},\ and\ \bibinfo {author}
  {\bibfnamefont {H.}~\bibnamefont {Montagu}},\ }\bibfield  {title} {\bibinfo
  {title} {{Formation of a non-magnetic, odd-denominator fractional quantized
  conductance in a quasi-one-dimensional electron system}},\ }\href
  {https://doi.org/10.1063/1.5121147} {\bibfield  {journal} {\bibinfo
  {journal} {Applied Physics Letters}\ }\textbf {\bibinfo {volume} {115}},\
  \bibinfo {pages} {123104} (\bibinfo {year} {2019}{\natexlab{b}})}\BibitemShut
  {NoStop}%
\bibitem [{\citenamefont {Gonze}\ \emph {et~al.}(2002)\citenamefont {Gonze},
  \citenamefont {Beuken}, \citenamefont {Caracas}, \citenamefont {Detraux},
  \citenamefont {Fuchs}, \citenamefont {Rignanese}, \citenamefont {Sindic},
  \citenamefont {Verstraete}, \citenamefont {Zerah}, \citenamefont {Jollet},
  \citenamefont {Torrent}, \citenamefont {Roy}, \citenamefont {Mikami},
  \citenamefont {Ghosez}, \citenamefont {Raty},\ and\ \citenamefont
  {Allan}}]{abinit}%
  \BibitemOpen
  \bibfield  {author} {\bibinfo {author} {\bibfnamefont {X.}~\bibnamefont
  {Gonze}}, \bibinfo {author} {\bibfnamefont {J.-M.}\ \bibnamefont {Beuken}},
  \bibinfo {author} {\bibfnamefont {R.}~\bibnamefont {Caracas}}, \bibinfo
  {author} {\bibfnamefont {F.}~\bibnamefont {Detraux}}, \bibinfo {author}
  {\bibfnamefont {M.}~\bibnamefont {Fuchs}}, \bibinfo {author} {\bibfnamefont
  {G.-M.}\ \bibnamefont {Rignanese}}, \bibinfo {author} {\bibfnamefont
  {L.}~\bibnamefont {Sindic}}, \bibinfo {author} {\bibfnamefont
  {M.}~\bibnamefont {Verstraete}}, \bibinfo {author} {\bibfnamefont
  {G.}~\bibnamefont {Zerah}}, \bibinfo {author} {\bibfnamefont
  {F.}~\bibnamefont {Jollet}}, \bibinfo {author} {\bibfnamefont
  {M.}~\bibnamefont {Torrent}}, \bibinfo {author} {\bibfnamefont
  {A.}~\bibnamefont {Roy}}, \bibinfo {author} {\bibfnamefont {M.}~\bibnamefont
  {Mikami}}, \bibinfo {author} {\bibfnamefont {P.}~\bibnamefont {Ghosez}},
  \bibinfo {author} {\bibfnamefont {J.-Y.}\ \bibnamefont {Raty}},\ and\
  \bibinfo {author} {\bibfnamefont {D.}~\bibnamefont {Allan}},\ }\bibfield
  {title} {\bibinfo {title} {First-principles computation of material
  properties: the abinit software project},\ }\href
  {https://doi.org/https://doi.org/10.1016/S0927-0256(02)00325-7} {\bibfield
  {journal} {\bibinfo  {journal} {Computational Materials Science}\ }\textbf
  {\bibinfo {volume} {25}},\ \bibinfo {pages} {478} (\bibinfo {year}
  {2002})}\BibitemShut {NoStop}%
\bibitem [{\citenamefont {Szabo}\ and\ \citenamefont {Ostlund}(1996)}]{Szabo}%
  \BibitemOpen
  \bibfield  {author} {\bibinfo {author} {\bibfnamefont {A.}~\bibnamefont
  {Szabo}}\ and\ \bibinfo {author} {\bibfnamefont {N.~S.}\ \bibnamefont
  {Ostlund}},\ }\href@noop {} {\emph {\bibinfo {title} {Modern Quantum
  Chemistry: Introduction to Advanced Electronic Structure Theory}}}\ (\bibinfo
   {publisher} {Dover Publications},\ \bibinfo {address} {Mineola, New York},\
  \bibinfo {year} {1996})\BibitemShut {NoStop}%
\bibitem [{\citenamefont {Dresselhaus}\ \emph {et~al.}(2007)\citenamefont
  {Dresselhaus}, \citenamefont {Dresselhaus},\ and\ \citenamefont
  {Jorio}}]{dresselhaus2007group}%
  \BibitemOpen
  \bibfield  {author} {\bibinfo {author} {\bibfnamefont {M.~S.}\ \bibnamefont
  {Dresselhaus}}, \bibinfo {author} {\bibfnamefont {G.}~\bibnamefont
  {Dresselhaus}},\ and\ \bibinfo {author} {\bibfnamefont {A.}~\bibnamefont
  {Jorio}},\ }\href@noop {} {\emph {\bibinfo {title} {Group theory: application
  to the physics of condensed matter}}}\ (\bibinfo  {publisher} {Springer
  Science \& Business Media},\ \bibinfo {address} {Berlin Heidelberg},\
  \bibinfo {year} {2007})\BibitemShut {NoStop}%
\bibitem [{\citenamefont {Clementi}\ \emph {et~al.}(1967)\citenamefont
  {Clementi}, \citenamefont {Raimondi},\ and\ \citenamefont
  {Reinhardt}}]{Clementi1967}%
  \BibitemOpen
  \bibfield  {author} {\bibinfo {author} {\bibfnamefont {E.}~\bibnamefont
  {Clementi}}, \bibinfo {author} {\bibfnamefont {D.}~\bibnamefont {Raimondi}},\
  and\ \bibinfo {author} {\bibfnamefont {W.~P.}\ \bibnamefont {Reinhardt}},\
  }\bibfield  {title} {\bibinfo {title} {{Atomic screening constants from SCF
  functions. II. Atoms with 37 to 86 electrons}},\ }\href@noop {} {\bibfield
  {journal} {\bibinfo  {journal} {The Journal of chemical physics}\ }\textbf
  {\bibinfo {volume} {47}},\ \bibinfo {pages} {1300} (\bibinfo {year}
  {1967})}\BibitemShut {NoStop}%
\end{thebibliography}%

\end{document}